\documentclass{article}

\usepackage{import}
\usepackage{booktabs}
\usepackage{babel}
\usepackage[x11names,table]{xcolor}
\usepackage[a4paper,top=3.25cm,bottom=3.25cm,left=3.25cm,right=3.25cm,marginparwidth=1.75cm]{geometry}
\usepackage{amsmath}
\usepackage{graphicx}
\usepackage{setspace}
\usepackage{makecell}
\usepackage{array,multirow}
\usepackage{subcaption}
\usepackage{url}
\usepackage[hidelinks]{hyperref}
\usepackage{cleveref}
\usepackage{algorithm}
\usepackage{algpseudocode}
\usepackage{lscape} 
\usepackage{makecell}
\usepackage{natbib}
\usepackage{microtype}

\newcolumntype{C}[1]{>{\centering\arraybackslash}p{#1}}

\title{Simple Macroeconomic Forecast Distributions for the G7 Economies\thanks{We thank the associate editor, two anonymous referees, Florian Huber, as well as workshop and conference participants at the German Council of Economic Experts,  the University of Salzburg, the University of Kiel, the IWH Workshop on Forecasting in Times of Structural Change and Uncertainty, the HKMetrics Workshop, the MathSEE Symposium, Statistische Woche 2024, the Annual Conference of the IAAE 2025 and the Vienna Workshop on Economic Forecasting 2025 for helpful comments.\\ 
F. Becker und M. Schienle gratefully acknowledge funding by the Helmholtz Association grant COCAP (KA1-Co-10). M. Schienle also thanks the Klaus Tschira Foundation.}}
\author{Friederike Becker$^1$, Fabian Kr\"uger$^1$ and Melanie Schienle$^{1,2}$\\[.4cm] $^1$ Karlsruhe Institute of Technology, Institute of Statistics\\ 
$^2$ Heidelberg Institute for Theoretical Studies}
\begin{document}
\onehalfspacing

\maketitle

\begin{abstract}
We present a simple method for predicting the distribution of output growth and inflation in the G7 economies. The method is based on point forecasts published by the International Monetary Fund (IMF), as well as robust statistics from the empirical distribution of the IMF's past forecast errors while imposing coherence of prediction intervals across horizons. We show that the technique  yields calibrated prediction intervals and performs similar to, or better than, more complex time series models in terms of statistical loss functions. We provide a simple website with graphical illustrations of our forecasts, as well as time-stamped data files that document their real-time character.
\end{abstract}
\newpage

\section{Introduction}

\label{sec:intro}

Macroeconomic forecasts feature prominently in economic policy debates, and outlooks are often compared in the cross-section of economically similar countries across the globe. While the academic literature on macroeconomic forecasting has long appreciated the importance of measuring and communicating forecast uncertainty (as represented, e.g., by distributions and prediction intervals), the broader policy debate is still dominated by point forecasts. This may partly be due to the fact that for many countries including the Group of 7 economies (henceforth G7, including Canada, France, Germany, Italy, Japan, the United Kingdom and the United States), real-time forecast distributions are not readily available. In case of the G7 and recent editions of macroeconomic forecasting reports issued by their central banks, three of the seven reports did not explicitly quantify uncertainty. Table \ref{tab:cbreports} and Section \ref{sec:cb_reports_detailed} of the supplement provide details. Among the many economic organizations that issue forecasts, central banks take a leading role in terms of both societal impact and technical sophistication. If anything, we thus expect their coverage of forecast uncertainty to be more advanced than that of other economic organizations. Hence, quantitative information on macroeconomic forecast uncertainty is often unavailable, even for well-informed readers who invest the time to consult rather specialized documents.  

Motivated by the discrepancy between macroeconomic forecasting in practice versus academia, we develop simple macroeconomic forecast distributions for growth and inflation in the G7 countries. These distributions are based on two main ingredients: Point forecasts provided by the International Monetary Fund (IMF), as well as appropriate quantiles from the empirical distribution of the IMF's past forecast errors that we adjust for coherent prediction intervals across horizons. Our approach is deliberately simple and fully automatic, transparent, and based on publicly available data only. All data and forecast intervals, as well as further replication materials are linked to in a public GitHub repository \citep{macropigit}.
The repository additionally contains forecasts for an ongoing and preregistered prospective exploration, with time stamps documenting their real-time character, and links to a graphical dashboard showing current prediction intervals (see Section \ref{sec:shiny}).

Transparent measurement and communication of forecast uncertainty is particularly important in economics, where forecasts are regularly used to motivate far-reaching policy decisions. In the last decades, various central bank related research initiatives have promoted probabilistic forecasting in macroeconomics. While these initiatives have inspired academic research on the topic, their impact on central bank decisions like interest rate setting has often been unclear \citep[c.f.][]{ConradEnders2024}. For the U.S., the Survey of Professional Forecasters \citep[SPF;][]{CroushoreStark2019} managed by the Federal Reserve Bank of Philadelphia has published probabilistic forecasts since 1968. For Europe, the ECB has maintained a similar survey since 1999 \citep{BowlesEtAl2007}. Both forecast surveys are made available in real time, following a clear release calendar. Probabilistic forecasts are available in the form of `histograms', i.e., participants' subjective probabilities for various outcome ranges (such as GDP growth between 0 and 2 percent). Given that the SPF and ECB-SPF collect forecasts (i.e., numbers) rather than forecasting methods (i.e., program code and data), the statistical techniques or judgmental components underlying the histogram-type forecasts are not known. Moreover, it is well known that the format of survey questionnaires as well as characteristics of forecasters influence the obtained results \citep[e.g.][]{GlasHartmann2022,Pavlova2024}.
By contrast, our forecast distributions are based on specified data inputs (IMF point forecasts, historical realizations data) and statistical techniques. Furthermore, we use a different (and arguably simpler) representation of forecast uncertainty via prediction intervals, and cover the G7 economies.
\renewcommand{\arraystretch}{1.2}
\begin{table}
\begin{center}
\begin{tabular}{l l c}
 \textbf{Country} &  \textbf{Name of Central Bank} \hspace*{5mm} &  \textbf{uncertainty quantified}\\[0.5em]
 \toprule \\[-1em]
 Canada & Bank of Canada & no \\  
 France & Banque de France & no \\ 
 Germany & Deutsche Bundesbank & no \\ 
 Italy & Banca d'Italia & yes \\ 
 Japan & Bank of Japan & somewhat \\ 
 United Kingdom \hspace*{5mm}& Bank of England & yes \\ 
 United States & Federal Reserve & yes    \\\bottomrule
 \end{tabular}
\caption{State of uncertainty quantification for macroeconomic forecasts that are issued by the central banks of the G7 countries. For each central bank, we consulted the latest ``flagship'' report as of April 2025 containing forecasts for inflation and real GDP growth, (often titled ``Macroeconomic Projections'', ``Economic Outlook'' or ``Monetary Policy Report'', depending on availability) and checked whether that document's centrally communicated forecasts are accompanied by an explicit quantification of uncertainty. For more details on the reports and our protocol for extracting the information contained therein, see Section \ref{sec:cb_reports_detailed} in the supplement.}
\label{tab:cbreports}
\end{center}
\end{table} 

Based on work by \cite{AdamsEtAl2021}, a recent project by the \cite{fed} provides probabilistic forecasts of real GDP growth, inflation and unemployment in the U.S. While similar in spirit, our forecast distributions are based on public data (rather than proprietary survey forecast data), and cover all G7 economies. In line with the IMF's World Economic Outlook, our forecasts are biannual (rather than monthly). Furthermore, our methodology is somewhat simpler, in that we do not attempt to predict the distribution of forecast errors by means of additional variables. \cite{schick2024} incorporates SPF forecasts into an AR-GARCH model to forecast quantiles of U.S. GDP growth. \cite{ReifschneiderTulip2019} describe prediction intervals presented by the U.S. Federal Reserve, based on root mean squared forecasting errors. The intervals' coverage level depends on the distribution of forecast errors, amounting to roughly 70 percent under a normal distribution. \cite{MPC2005} evaluate prediction intervals implied by the Bank of England's `fan charts', which are based on a skewed distribution reflecting the Bank's judgment. Compared to these parametric approaches, our proposed technique considers empirical quantiles of absolute forecast errors, thus avoiding restrictive functional form assumptions. The robustness of quantiles is an important advantage in turbulent times like the Covid-19 pandemic, whose extreme observations are handled plausibly and automatically by quantiles. By contrast, non-robust estimation methods based on squared errors are dominated by these observations. This leads to difficult questions about how to handle extreme observations in practice, with many studies resorting to ad-hoc choices of the sample period. See \cite{LenzaPrimiceri2022} and \citet[Section 5]{KnueppelEtAl2023} for further discussion. 

Complementary to our focus on forecast distributions at longer horizons, \cite{KronenbergEtAl2023} provide a detailed platform for visualizing and downloading point forecasts. They study GDP growth at short horizons (current and next quarter), covering various countries and modeling approaches. 

Finally, our paper relates to a growing body of academic literature on how to construct forecast distributions of macroeconomic variables. Our proposed approach intends to be as simple as possible (in terms of data requirements and statistical techniques), subject to being reasonably competitive in terms of statistical performance. Studies like \cite{Clements2010}, \cite{Krueger2017}, \cite{GanicsEtAl2023} and \cite{KruegerPlett2023} indicate that forecast distributions based on past point forecast errors perform similar to, or better than, subjective histogram-type forecasts as provided by the SPF and ECB-SPF. These findings motivate our use of the IMF WEO as an external point forecast, together with a suitable set of historical forecast errors. See \cite{QuEtAl2023} for recent evidence on the good performance of IMF point forecasts relative to private sector survey forecasts. The IMF point forecasts are survey based and may thus incorporate judgmental information that is hard to incorporate into formal time series models (e.g., information on fiscal or monetary policy announcements, or recent releases of weekly or monthly macroeconomic time series). For U.S. data, studies such as \cite{FaustWright2009} and \cite{KruegerEtAl2017} have found that judgmental information is most helpful at short forecast horizons (where survey forecasts of GDP growth and inflation are hard to beat), while being less effective at longer horizons. By using the IMF point forecast, our method exploits the potential benefits of judgmental information in terms of point forecasting. At the same time, and as motivated above, our assessment of forecast uncertainty is purely statistical. Compared to studies that create `self-contained' forecast distributions based on multivariate time series data \citep[see e.g.][and the references therein]{ClarkEtAl2024}, our use of the IMF WEO as an external point forecast greatly reduces data requirements and modeling complexity. For our purposes, this simplification outweighs the inherent benefits of producing a self-contained forecast. More broadly, the principle of constructing forecast distributions based on a history of point forecasts and associated realizations has proven successful in many empirical contexts across the disciplines. Similar to our main approach, \cite{TulipWallace2012} explore the use of quantiles from the history of absolute errors to quantify uncertainty for the Reserve Bank of Australia's forecasts. In the context of high-frequency electricity spot price data, \cite{KathZiel2021} similarly include a model that constructs forecast intervals based on quantiles of absolute point forecast errors. \cite{WalzEtAl2024} and \cite{AB2023} provide further discussion in the context of meteorology and machine learning, respectively. While we focus on forecast errors constructed from IMF point forecasts, the methodological questions we discuss (in particular, the use of absolute versus raw forecast errors, and the concrete specification  of window method for selecting the relevant training data) arise similarly when using any other source of point forecasts. These other sources include statistical or machine learning methods for which point forecasts are often more easily accessible via software packages than quantile or distribution forecasts. This paper hence offers a framework for designing error-based forecast distributions in empirical macroeconomics. 

The remainder of this paper is organized as follows. Section \ref{sec:methods} contains methodology for computing empirical quantiles in the current setup. In particular, we discuss techniques for ensuring that forecasts are coherent across multiple horizons, as well as assumptions about the (a)symmetry of forecast errors. Section \ref{sec:appres} describes relevant benchmark methods and presents empirical results on the forecasting performance of the proposed method. In our empirical analysis, we show that the method yields calibrated prediction intervals (with coverage close to its nominal level), and overall performs similarly to or better than the benchmarks in terms of the interval score, a statistical loss function for prediction intervals. Section \ref{sec:conc} concludes. The supplement contains details on Table \ref{tab:cbreports}, as well as further empirical results and robustness checks. Replication materials for the paper are available through \cite{macropigit}.

\section{Methods}\label{sec:methods}
\subsection{Constructing Prediction Intervals}
\label{sec:constructing}
This paper aims to provide a calibrated probabilistic assessment 
of macroeconomic indicators in an accessible manner: First, the format of the probabilistic forecasts should be easily understandable to facilitate their communication and second, the methodology should be simple, to make the forecast generation transparent and readily reproducible.  

We use prediction intervals as a method for capturing uncertainty. This format has been shown to be successful in a meteorological  \citep{probcast09,raftery_prob} and public health \citep{bracher2021pre, cramer2022evaluation} context. While conveying less information than, for instance, a full probability distribution, they are easier to use and understand for non-statisticians: For a given confidence level $\tau \in (0, 1)$, a prediction interval is represented by only two numbers $u^{\tau} > l^{\tau}$, the upper and lower endpoints of the interval.  

We use an existing series of point forecasts for macroeconomic indicators, and construct a prediction interval for a given level $\tau$ around the point forecast $\hat{y}$ in the following way: 
\begin{align}\label{eq:cigeneral}
\begin{split}
	u^{\tau} &= \hat{y} + c^{u, \tau}\\
	l^{\tau} &= \hat{y} + c^{l,\tau},
\end{split}
\end{align}
 with $c^{u, \tau} > c^{l, \tau}$. Note that $c^{l, \tau}$ can, but does not have to, be negative. To get a finer representation of the predictive distribution, one can construct these intervals for multiple levels of $\tau$. Given $\tau$, we obtain values for the upper and lower prediction bands $c^{u, \tau} \text{ and } c^{l, \tau}$ from past data on forecasts and realizations, according to the following methodology.
 
\begin{figure}
\centering\includegraphics[width=.7\linewidth]{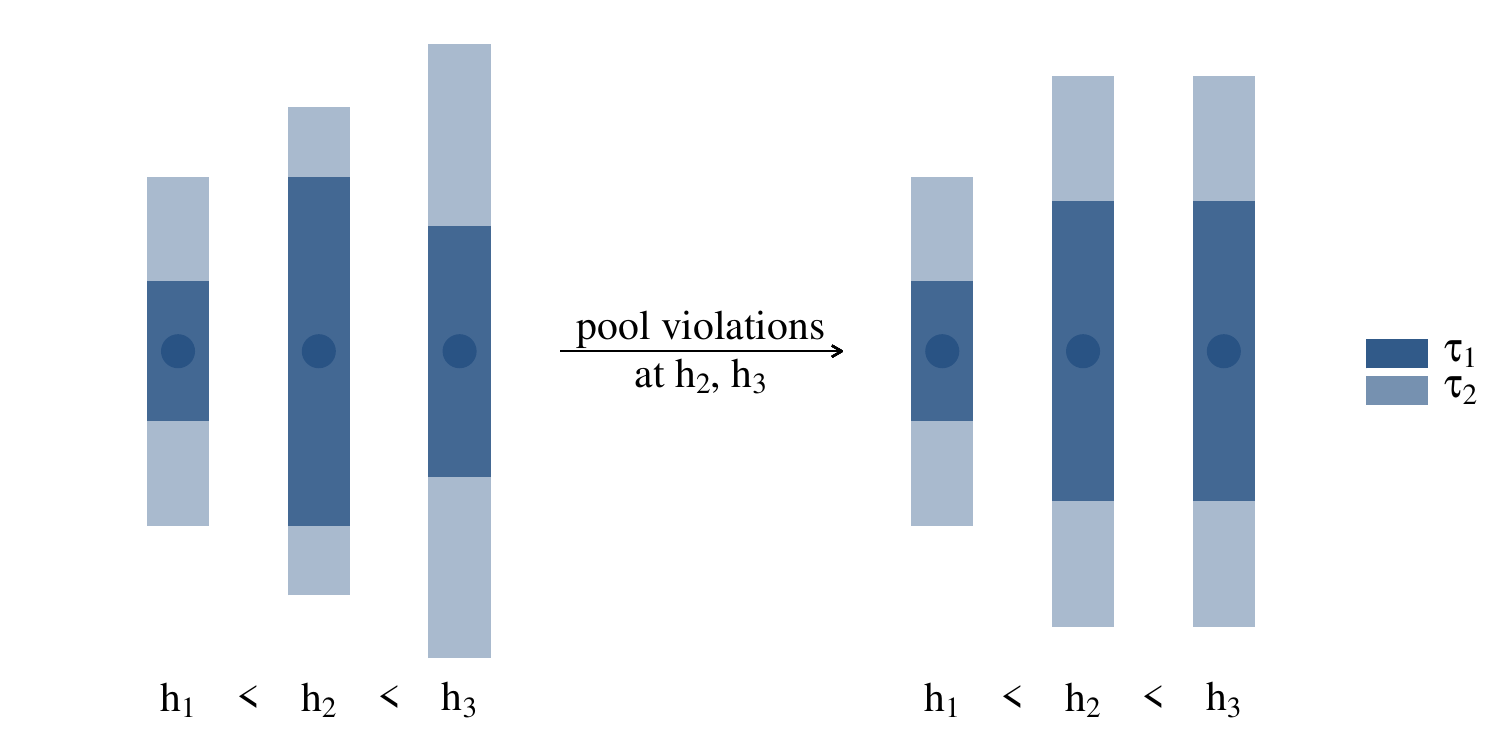}
\caption{Illustration of the PAVA-type correction procedure of Algorithm \ref{alg:pava} used in cases where the monotonicity constraint is not upheld across horizons. The figure shows prediction intervals for two confidence levels $\tau_1$ (dark shaded area) and $\tau_2$ (light shaded area), with $\tau_1 < \tau_2$, and three horizons $h_1 < h_2 < h_3$. The prediction intervals for level $\tau_1$ violate the constraint \eqref{eq:horizonordering} at horizons $h_2$ and $h_3$. As a result, prediction intervals at both levels $\tau_1$ and $\tau_2$ are pooled at $h_2$ and $h_3$, averaged, and thereafter assigned to both horizons. For illustration purposes, all forecasts were centered at the same value.\label{fig:ci_illustrations}} 
\end{figure}
For a given country and forecast target, suppose we have access to a past series of forecasts $\{\hat{y}_{t, h}\}_{t=1,...,T}$, as well as to the corresponding realized values $\{y_{t}\}_{t=1,...,T}$. The series of forecasts must have both sufficient history and still be ongoing. The target year is indexed by $t$ and the forecast horizon by $h$. The latter denotes the time difference between the date the forecast is made (the ``forecast origin'') and the end of year $t$, when the quantity realizes.

Given these forecast-observation pairs, we construct sets 
\begin{equation}\label{eq:errorset}
\mathcal{E}^{abs}_{t, h} = \{\hat{e}^{abs}_{t^*, h}  | t-R \leq t^* < t \}
\end{equation}
based on the \textit{absolute} forecast errors 
\begin{equation}\label{eq:abserrors}
\hat{e}^{abs}_{t,h} = |y_{t} - \hat{y}_{t, h}|.
\end{equation}
For a discussion of why we choose absolute forecast errors rather than raw error values, we refer to Section \ref{sec:errormethod}. In cases where $h$ is larger than one year and the directly preceding target year(s) have not yet completed, the size of the set \eqref{eq:errorset} is kept fixed at $R$ by incorporating the appropriate number of previous years. Apart from practical constraints imposed by a limited sample size, the choice of $R$ reflects a trade-off between stability (which suggests a large value of $R$) and emphasizing recent observations (which suggests a small value of $R$); see e.g. \citet[Section 3]{GneitingEtAl2005}. 

Given \eqref{eq:errorset}, we obtain the lower $l$ and upper $u$ endpoints of a prediction interval of desired level $\tau$ via adding and subtracting the empirical $\tau-$quantile of \eqref{eq:errorset} to the current point forecast:
\begin{align}\label{eq:abs_ci}
\begin{split}
	u^{\tau}_{t,h} &= \hat{y}_{t, h} + q^{\tau}\left(\mathcal{E}^{abs}_{t, h}\right)\\
	l^{\tau}_{t,h} &= \hat{y}_{t, h} - q^{\tau}\left(\mathcal{E}^{abs}_{t, h}\right), 
\end{split}
\end{align}
where $q^{\tau}$ refers to the empirical quantile function. Due to the monotonicity of the empirical quantile function $q^{\tau}$ in $\tau$ , quantile crossing is avoided by construction across different levels of $\tau$. Note also that we obtain a symmetric interval since here $c^{l,\tau} = - c^{u,\tau}$.

As discussed by \cite{hyndman1996}, several methods for computing empirical quantiles are implemented in \textsf{R} \citep{R2023} and other statistical programming environments.  
As detailed in Section \ref{sec:tuning}, we will give preference to combinations of an empirical quantile method and window length $R$ that, subject to data availability and the chosen confidence levels, allow extraction of quantiles directly from the order statistics of the forecast errors in $\mathcal{E}_{t,h}$ and thus do not rely on interpolation between observed values. This choice is motivated by simplicity, in that interval endpoints that stem directly from observed forecast errors seem easier to understand.

Many economic organizations (such as the IMF) simultaneously issue forecasts for multiple future years. Here we impose the assumption that the length of prediction intervals should not decrease with the forecast horizon.
This assumption is intuitively appealing, and also aligns with theoretical notions of optimal forecasting in stationary time series models \citep[see e.g.][]{PattonTimmermann2011, KruegerPlett2023}. In general, a sufficient condition for the assumption is that 
\begin{align}\label{eq:horizonordering}
\begin{split}
 c^{u,\tau}_{t,h} &\leq  c^{u,\tau}_{t,h'}, \textrm{ for all } h<h', \text{ and}\\ 
 c^{l,\tau}_{t,h} &\geq  c^{l,\tau}_{t,h'}, \textrm{ for all } h<h';
\end{split}
\end{align}
for the case of our symmetric method (with $c^{l, \tau} = - c^{u, \tau}$), it is sufficient to check either one of the inequalities. We impose this restriction on the values $c^{l, \tau}, c^{u, \tau}$ rather than directly on the length of the prediction interval in order to simplify the adjustment mechanism in case of violations.  In particular, if monotonicity is violated, we enforce it via the pool-adjacent-violators (PAVA) type reordering outlined in Algorithm \ref{alg:pava}. See e.g. \cite{deLeeuwEtAl2009} for background on the PAVA algorithm. In short, the procedure amounts to iteratively merging predictive intervals at all confidence levels in case of violations, until the condition in \eqref{eq:horizonordering} is upheld across all horizons. Since the PAVA-type reordering is applied to all considered quantile levels if there is a violation of \eqref{eq:horizonordering} at one level, the reordering does not cause any quantile crossing.
\begin{algorithm}[t]
\begin{algorithmic}
\State The initial solution is  $c^{u,\tau}_{t,r}(0) := q^{\tau}\left(\mathcal{E}^{abs}_{t, r}\right)$ and $c^{l,\tau}_{t,r}(0) := - q^{\tau}\left(\mathcal{E}^{abs}_{t, r}\right)$, where $q$ is the chosen empirical quantile function. The index for the blocks is $r = 1, . . . , B$ where $B$ is initially the number of forecast horizons.
\While{there exists $r, \tau'$ such that:  $c^{u,\tau'}_{t,r}(k) > c^{u,\tau'}_{t,r+1}(k)$} 
	\State For all values of $\tau$: Merge $c^{u,\tau}_{t}(k)-$values from blocks $r$ and $r+1$ into block $r$ and average. Merge $c^{l,\tau}_{t}(k)-$values from blocks $r$ and $r+1$ into block $r$ and average.  
 \EndWhile
\end{algorithmic}
\caption{PAVA-type algorithm for symmetric prediction intervals.\label{alg:pava}}
\end{algorithm}

\subsection{Absolute versus Raw Errors}\label{sec:errormethod}
As noted previously, we rely on absolute errors to construct prediction intervals. We next compare absolute errors to raw, ``directional'' errors given by
\begin{equation}
\hat{e}^d_{t,h} = y_{t} - \hat{y}_{t, h}.
\end{equation}
The relevant sets of errors $\mathcal{E}^d_{t,h}$ are constructed analogously and the prediction interval endpoints can be calculated as 
\begin{align*}
	u^{\tau}_{t,h} &= \hat{y}_{t, h} + q^{\frac{1+\tau}{2}}\left(\mathcal{E}_{t, h}^d\right)\\
	l^{\tau}_{t,h} &= \hat{y}_{t, h} + q^{\frac{1-\tau}{2}}\left(\mathcal{E}_{t, h}^d\right).
\end{align*}
A major drawback of this method is as follows: Especially when few past observations of the series are available, and with smaller confidence levels such as $\tau=0.5$, it is possible that either the $\frac{1-\tau}{2}$-quantile is positive or the $\frac{1+\tau}{2}$-quantile is negative. In particular, this means that the point forecast would not be contained in the prediction interval that is constructed ``around'' it. This type of one-sided behavior seems practically relevant in the empirical setup we consider below.

On the other hand, prediction intervals based on ``directional'' errors are entirely unrestricted, whereas the use of absolute errors \eqref{eq:abs_ci} comes at the cost of implicitly assuming that the distribution of forecast errors is symmetric around zero \citep[c.f.][]{Breth1982}. This is a rather stringent assumption. It implies that the point forecast corresponds to the median functional of the forecast distribution. Furthermore, symmetry implies that the median functional equals the mean functional. In this sense, the assumption of symmetric forecast errors sidesteps the debate about which functional of the forecast distribution is being addressed by a given point forecast \citep[e.g.][]{ElliottEtAl2005,Manski2018}. From a statistical perspective, the assumption that forecast errors are symmetric around zero regularizes the estimate of the forecast error distribution. This form of regularization seems beneficial if its implied assumptions are at least approximately correct, and if the sample size is small, as is the case in our application. In the related context of estimating a conditional (rather than unconditional) distribution of forecast errors, \cite{KruegerPlett2023} find that imposing symmetry around zero is helpful in terms of forecasting performance. Furthermore, various types of prediction intervals based on absolute forecast errors have been studied in the literature on conformal prediction; see e.g. \citet{ShaferVovk2008} and \cite{KathZiel2021}.

Table \ref{tab:proscons} summarizes our general considerations for deciding between absolute versus directional errors. Section \ref{sec:dir_illustrative} in the supplement also presents illustrative evidence of  empirical advantages of absolute versus directional errors for our specific small sample setup. In view of these arguments, we opt to use absolute errors as our default option. Of course, this choice reflects a weighting of the different criteria mentioned in Table \ref{tab:proscons}, which is necessarily subjective. 

\begin{center}
	\begin{table}[htbp]
		\begin{tabular}{p{0.05\linewidth}  p{0.36\linewidth} p{0.025\linewidth} p{0.05\linewidth}  p{0.36\linewidth} }
			
			\multicolumn{2}{c}{\large{\textbf{Absolute Errors}}}& & \multicolumn{2}{c}{\large{\textbf{Directional Errors}}}\\[1.8em] 
			\multicolumn{5}{l}{\small{\textbf{Shape of prediction interval}} }\\
			\midrule \multicolumn{5}{l}{} \\[-0.5em]
			$+$ & Prediction intervals are centered around the existing point forecast. This may be seen as intuitive. & 
			& $-$& Prediction intervals are not centered around the point forecast and may not even contain it. Especially for intervals with a high nominal level of confidence, this may be seen as unintuitive. \\
			&&&&\\
			\multicolumn{5}{l}{\small{\textbf{Assumptions}} }\\
			\midrule
			\multicolumn{5}{l}{} \\[-0.5em]
			$-$ & True distribution of forecast errors must be symmetric around zero (otherwise, use of absolute errors is suboptimal in large samples). Accordingly, the external point forecast is interpreted as referring to the median functional, which coincides with the mean functional in this case. & 
			& $+$& No distributional assumption on forecast errors required. Accordingly, no assumption regarding the interpretation of the external point forecast.\\
			&&&&\\
			$+$ & Implicit symmetry assumption on distribution of forecast errors regularizes quantile estimation, which can be beneficial in small samples \citep[c.f.][]{Breth1982}.   & 
			& $-$& No regularization, and quantile estimation can be challenging in small samples.\\ 
			&&&&\\
			\multicolumn{5}{l}{\small{\textbf{In our application (Section \ref{sec:appres}): Scores and calibration}} }\\
			\midrule
			\multicolumn{5}{l}{} \\[-0.5em]
			$+$ & Coverage rates are close to nominal. & 
			& $-$& Interval scores are mostly similar to absolute error method, but coverage rates are often substantially below nominal rates.\\ \bottomrule
		\end{tabular}
		
		\caption{Advantages and disadvantages of using either absolute error calculation (left column) or directional error calculation (right column) for constructing forecast intervals. Advantageous points are indicated with $+$, disadvantageous points with $-$.\label{tab:proscons}}

	\end{table}
\end{center}

\subsection{Assessing Forecast Accuracy}
\label{sec:performance_measures}
In assessing the quality of our forecasts, we follow the paradigm to `maximize sharpness subject to calibration' postulated by \cite{GneitingEtAl2007}. Sharpness means that the forecast distribution should be as narrow as possible. Calibration means that the forecast distribution should be coherent with observed outcomes. In practice, there is a clear trade-off between both objectives. For example, a forecast distribution with all probability mass in one point (indicating no uncertainty) would be perfect in terms of sharpness, but likely poor in terms of calibration. In order to assess the calibration of probabilistic forecasts in an interval format, a simple method is to compare the empirical and nominal interval coverage rates. That is, a $\tau\%$ coverage interval should over time cover roughly $\tau\%$ of observations. As argued by \cite{raftery_prob}, forecast users tend to view calibration as a crucial feature of a trustworthy forecast in practice. A proper scoring rule can then be used to evaluate the forecast sharpness in conjunction with its calibration properties. Briefly, a scoring rule is called proper if it encourages forecasters to state what they think is the correct prediction \citep{Winkler1996, GneitingRaftery2007}. Given the format of our forecast, the interval score \citep{GneitingRaftery2007} is a natural choice of scoring rule:
\begin{equation} \label{eq:interval_score}
\text{IS}_{\tau}(F, y) = (u-l) + \frac{2}{1-\tau}(l - y)1(y < l) + \frac{2}{1-\tau}(y - u)1(y > u),
\end{equation}
for a given confidence level $\tau$ and the corresponding prediction interval $[l,u]$ implied by the forecast distribution $F$. Here we consider scores in negative orientation, i.e., smaller scores correspond to better forecasts. The first summand of Equation (\ref{eq:interval_score}) can be interpreted as a penalty for dispersion (i.e., lack of sharpness) of the forecast distribution, corresponding to wide prediction intervals, such that $u-l$ is large. The second summand can be seen as a penalty for overprediction, such that the outcome $y$ is smaller than the interval's lower endpoint $l$. Conversely, the third summand of (\ref{eq:interval_score}) represents a penalty for underprediction. Given that (weighted) sums of proper scoring rules are again proper \citep{GneitingRaftery2007}, we can add scores for different levels of $\tau = \tau_1, \ldots, \tau_K$ in order to obtain a summary measure of forecast performance. In the following, we consider a weighted sum with weights $w_1,...,w_K$ as proposed by \cite{bracher2020}. Specifically, we set
$K = 2$, $\tau_1 = 0.5$ and $\tau_2 = 0.8$, and thereby $w_1 = \frac{1-\tau_1}{2}= 0.25$ and $w_2 = \frac{1-\tau_2}{2}= 0.1$. Since we focus on generating prediction intervals, we do not incorporate the point (median) prediction into the interval score. The latter is externally provided by the IMF in our empirical setup.  

\section{Empirical Results for the G7 Economies}\label{sec:appres}
In this section, we apply the methods described in Section \ref{sec:methods} to the IMF's World Economic Outlook (WEO) forecasts.
\subsection{Data and Forecast Targets}\label{sec:data}
The WEO's merit relative to other forecast sources has been established in studies such as \cite{timmermann2008} and \cite{QuEtAl2023}, which is why we consider it a promising candidate for this analysis. The WEO issues biannual forecasts for multiple (up to six) years into the future, for several variables including real GDP growth and inflation \citep{WEODB}. We focus on the latter two variables throughout this paper. The code that downloads and processes the current WEO dataset from the IMF website is available through our GitHub repository \citep{macropigit}.

For both the calculation of forecast errors and the subsequent evaluation of forecasts, we use the truth values that are included in the WEO data. As is typical for macroeconomic data, these are updated over time, due to relevant information becoming available retrospectively. For example, the best estimate of GDP in the year 2022 is typically different in spring 2023 (the first WEO estimate) than in fall 2023 (the second WEO estimate). While earlier studies of the WEO, such as \cite{timmermann2008}, choose different truth values depending on the forecast horizon, we opt for a consistent choice and mainly use the truth value that is published in the fall release following the year in question (e.g., fall 2023 for 2022). When constructing forecast intervals in spring, we however take a pragmatic approach and use the spring release's truth value for the directly preceding year, as the fall release for this year is not available until six months later. 

Our analysis focuses on the G7 economies and forecasts for the current and next year. We compute these forecasts at each biannual forecast date (spring and fall) covered by the IMF. This concise selection allows an accurate review and communication of results. We generally pool forecast evaluation results across the seven countries in order to increase the sample size. Of course, given the dependence of observations across countries, this step does not simply increase our effective sample size by seven.

We focus on the 80\% and 50\% prediction intervals. As argued by \cite{raftery_prob}, an 80\% interval satisfies the intuitive notion that a prediction interval should `typically' cover the realizing outcome, while often being reasonably short (i.e., sharp) in practice. The $50\%$ interval is additionally constructed to provide further characterization of the forecast distribution for interested users, and to serve as a further check of the validity of the methodology.

\subsection{Choice of Tuning Parameters}\label{sec:tuning}

The IMF data set covers forecasts and realizations for the period 1990-2023. We split the data into a training sample ranging from 1990-2012, and a hold-out sample ranging from 2013-2023. Note that next-year forecasts are only available for target year 1991 and onward, so that the training set is reduced by one observation for the two next-year horizons.

Our method contains two main tuning parameters to be chosen: The time window $R$ of past data used for quantile estimation, and the choice between absolute and directional forecast errors. Our choice of these parameters is guided by practical and conceptual concerns (see Section \ref{sec:methods} and Table \ref{tab:proscons}) and by empirical performance in the training data. In particular, we did not look at the hold-out data for choosing tuning parameters.

Table \ref{tab:dirvsabs} and Figure \ref{fig:rwlength} in the supplement summarize the training sample performance for the different combinations of the tuning parameters. Based on the performance in the training sample, and the considerations in Section \ref{sec:methods}, we use a rolling window of $R=11$ observations, the absolute error method, and the default empirical quantile method (\textsf{type = 7})\footnote{Note that for $R=11$ and the absolute error method, this is also equivalent to the inverse of the empirical distribution function, that is, \textsf{type = 1}. As we do not estimate quantiles at the tails of the distribution the arguments of \cite{hyndman1996} in favor of the \textsf{type = 8} method do not apply for our setting.} of the \textsf{quantile} function in \textsf{R}. Our training sample results indicate that shorter window lengths $R < 11$ perform worse in terms of the IS, see Figure \ref{fig:rwlength} in the supplement. The difference in performance is particularly pronounced for very short window lengths $4 \le R < 7$ and at longer forecast horizons. By contrast, the benefit of increasing $R$ from $8$ to $11$ is remarkably small in all empirical setups (absolute and directional errors, all forecast horizons). Hence the marginal benefit of increasing the sample size seems to vanish rapidly as $R$ increases. We exclude choices $R > 11$ since we cannot evaluate their performance given the time span of our training data, while an extension of the training data would have reduced the time span of our holdout sample, thus limiting our ability to assess out-of-sample performance. From a statistical perspective, using a short window length $R$ ensures adaptability to potential structural breaks \citep[c.f.][]{InoueEtAl2017}. We find that coverage rates on the training data are generally more favorable for the absolute than for the directional error method, sometimes on the order of about 10 percentage points (see Table \ref{tab:dirvsabs} in the supplement), while interval scores are similar. We additionally present illustrative insights from the training period in Section \ref{sec:dir_illustrative} in the supplement. Albeit attractive in principle, we argue that directional forecast errors sometimes display undesirable practical properties in our setup, and are thus not a good candidate to account for a potential skew in the distribution of forecast errors.

\begin{figure}
\centering
\includegraphics[width = 0.8\linewidth]{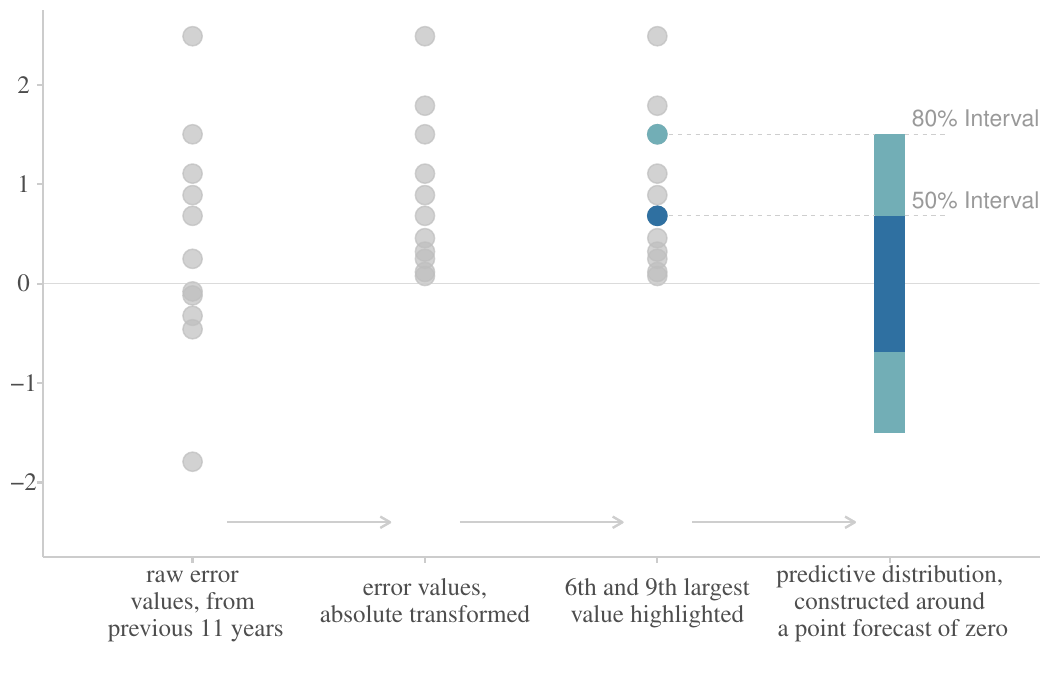}
\caption{Illustration of the process that we use to construct prediction intervals at the $50\%$ and $80\%$ confidence levels, using an existing point forecast and its history of forecast errors.}
\label{fig:illustration_quantileextraction}
\end{figure}

In addition to good training sample performance, the combination of $R = 11$ and the absolute error method is attractive in terms of computation and automated transparency of the algorithm facilitating communication: The endpoints of the $80\%$ prediction interval can simply be computed as plus or minus the $9$th largest absolute error in the $11$ years preceding the forecast date. Similarly, the endpoints of the $50\%$ interval can be computed from the $6$th largest absolute error. Figure \ref{fig:illustration_quantileextraction} illustrates this calculation.

\noindent

\subsection{Benchmark Models}\label{sec:bench}

In order to put the performance of our proposed method in perspective, we compare it to several benchmarks. For ease of presentation, we focus on three main benchmark models in our main empirical analysis that each represent one of the following broader model types: a) univariate or b) multivariate time series models yielding different point predictions, where prediction intervals are calculated either based on past performance forecast errors as in our approach or based on c) the respective time series model specification. For further variants of these benchmark models and respective comparisons we refer to Section \ref{sec:add_bench} below as well as Sections \ref{sec:details_bench} and \ref{sec:results_bench} of the supplement. In particular, Tables \ref{tab:extcis_currentyr} and \ref{tab:extcis_nextyr} of the supplement provide comprehensive results in terms of evaluation metrics.

All considered benchmarks are of (vector) autoregressive time series form, which is one of the most common choices in macroeconomic forecasting. For example, \cite{FaustWright2009} show that simple univariate autoregressions provide good forecasts of U.S. GDP growth when equipped with an appropriate information set, and \cite{CarrieroEtAl2024b} argue that vector autoregressive models with stochastic volatility compare favorably to nonlinear quantile regression methods for U.S. GDP growth, inflation and unemployment.
We fit the main benchmark models to quarterly data from the OECD for inflation and GDP growth (see Section \ref{sec:bmdata} in the supplement), using expanding windows of training data. In order to roughly match the timing of IMF forecasts, we equip the model-based spring forecasts with data up until the first quarter, and the model-based fall forecasts with data up until the third quarter. As we argue in Section \ref{sec:oecdtruth}, evaluation with respect to IMF data is still justified.

For the first main benchmark (denoted AR), we compute point predictions from an autoregressive model of order one, estimated via ordinary least squares. We construct a point forecast of the annual series as a weighted sum of quarterly forecasts and calculate prediction intervals from empirical quantiles of the model's past forecast errors, analogous to our proposed procedure for the IMF forecasts. In Section \ref{sec:add_bench} we also allow for a more general AR($p$) variant with data-driven choice of the lag length parameter $p$, and consider autoregressive models fitted to scarce annual IMF data.

For the second main benchmark (denoted BVAR), we consider a bivariate Bayesian vector autoregressive specification for inflation and GDP growth. As for the IMF forecasts (and for the AR model), we construct prediction intervals from empirical quantiles of the BVAR's past forecast errors. Our BVAR implementation is based on the \textsf{R} package \textsf{bvarsv} \citep{Krueger2015}. While the latter considers the flexible model by \cite{Primiceri2005} and \cite{delNegro2015}, we restrict the priors in our main benchmark to shut off time variation in the model's mean and covariance parameters, as in \citet[Section 5]{KnueppelEtAl2023}. We find that this simpler specification yields relatively stable parameter estimates for all countries and time periods. By contrast, \citeauthor{Primiceri2005}'s original priors that allow for time-varying parameters and stochastic volatility run into problems for some countries such as France and Great Britain after the extreme GDP growth rates observed at the onset of the Covid-19 pandemic.\footnote{This problem could possibly be resolved by adjusting the model's prior parameters to allow for an intermediate degree of flexibility, or by using a stochastic volatility specification that explicitly addresses Covid-19 outliers \citep[as in][]{CarrieroEtAl2024a}. However, the details of such an adjustment would create many additional degrees of freedom in our setup, given that the severity of the Covid-19 outliers differs across the G7 countries. We hence use a simpler specification with constant parameters for our main analysis.} Within the additional analyses discussed in Section \ref{sec:add_bench}, we also consider an approach that uses a stochastic volatility specification until 2019 and a constant volatility specification afterwards. This mixed variant entails a manual choice of regime switch. Even then, however, the performance of the model with constant parameters is largely similar with the exception of inflation at the next-year horizons where the mixed BVAR can outperform all methods and in particular our simple IMF-based approach.

For the third main benchmark (denoted BVAR-direct), we use the same constant parameter specification as in BVAR, but compute prediction intervals directly from the model's forecast distribution. That is, the forecast distribution is based on the model's parametric assumptions, rather than on the empirical distribution of the model's past point forecast errors. As is typical of Bayesian approaches, the BVAR-direct forecast distribution incorporates the effect of parameter estimation uncertainty. For completeness, Section \ref{sec:add_bench} also considers ``direct'' variants of the other benchmark specifications.

Besides variations of the main benchmark models, Section \ref{sec:add_bench} also contains further benchmark approaches such as models that use additional regressors and a simple ensemble.

\subsection{Forecast Performance on the Hold-Out Data}

\begin{figure}
  \begin{subfigure}{0.475\linewidth}
  \centering
    \includegraphics[width=\linewidth]{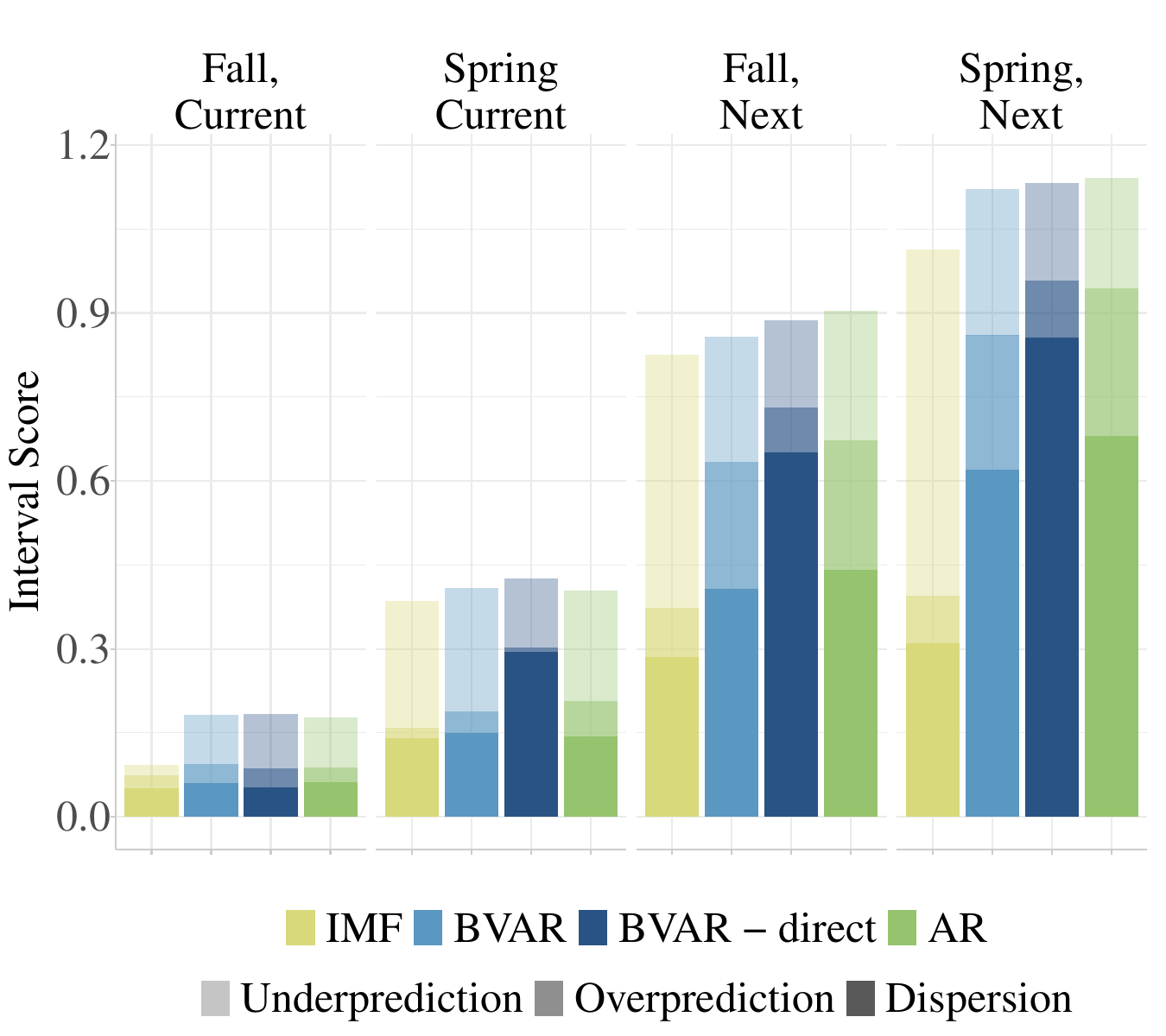}
    \caption{Inflation} \label{fig:ho_wis_pcpi_pch_absolute_rollingwindow}
  \end{subfigure}
  \hfill  
  \begin{subfigure}{0.475\linewidth}
  \centering
    \includegraphics[width=\linewidth]{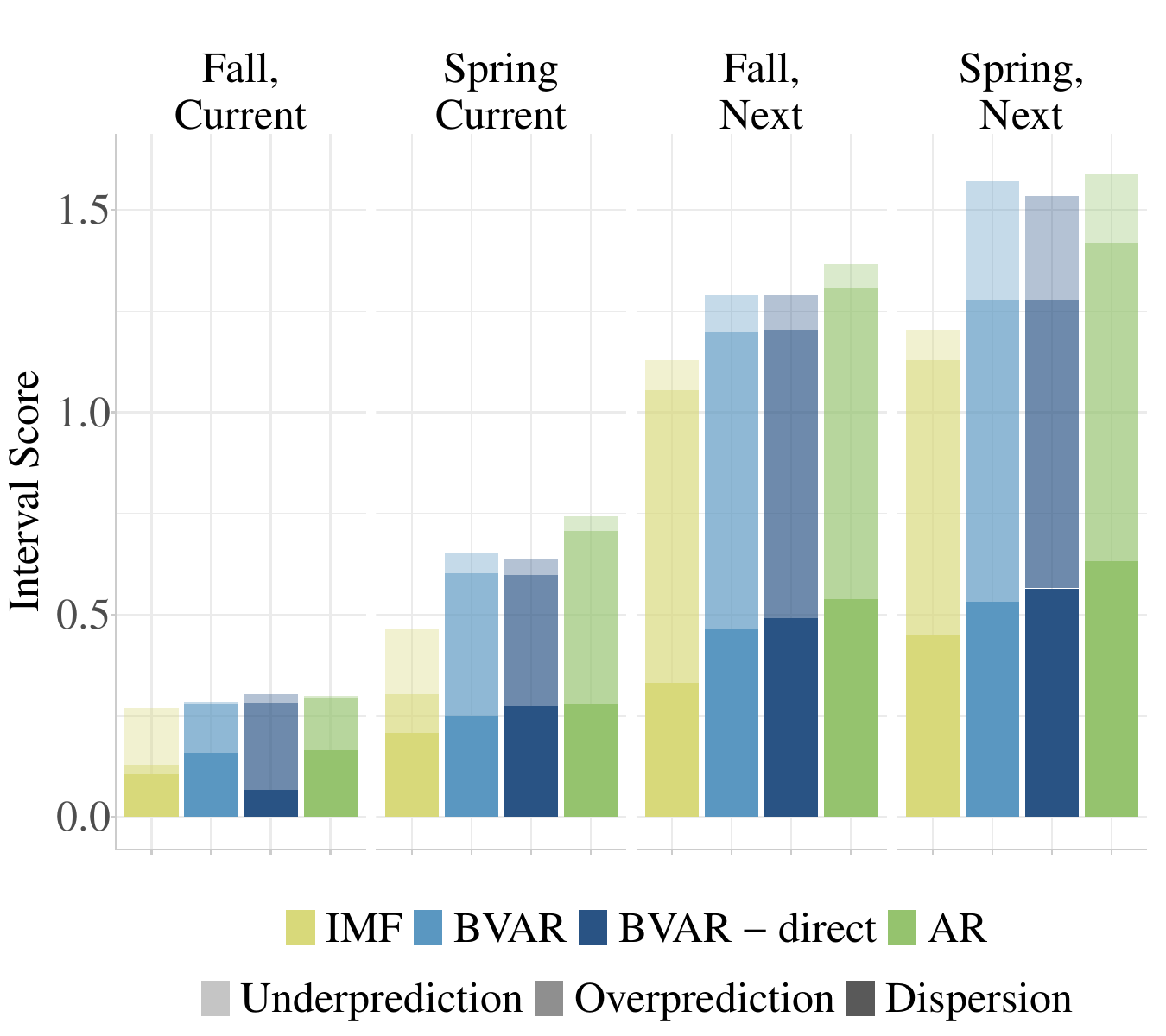}
    \caption{GDP growth} \label{fig:ho_wis_gdp_absolute_rollingwindow}
  \end{subfigure}
\caption{Interval scores (IS), separately for (a) inflation and (b) GDP growth, evaluation period 2013-2023. For each forecast horizon, the IS is shown side-by-side for the four methods. The IS is decomposed into underprediction, overprediction and dispersion, as indicated by the vertically stacked shaded areas. Scores are computed as a weighted sum of interval scores at the $50\%$ and $80\%$ level.} \label{fig:ho_wis_absolute}
\label{fig:ho_wis_absolute_rollingwindow}
\end{figure}

\begin{figure}
	\begin{subfigure}{\linewidth}
		\centering
		\includegraphics[width=1.01\linewidth]{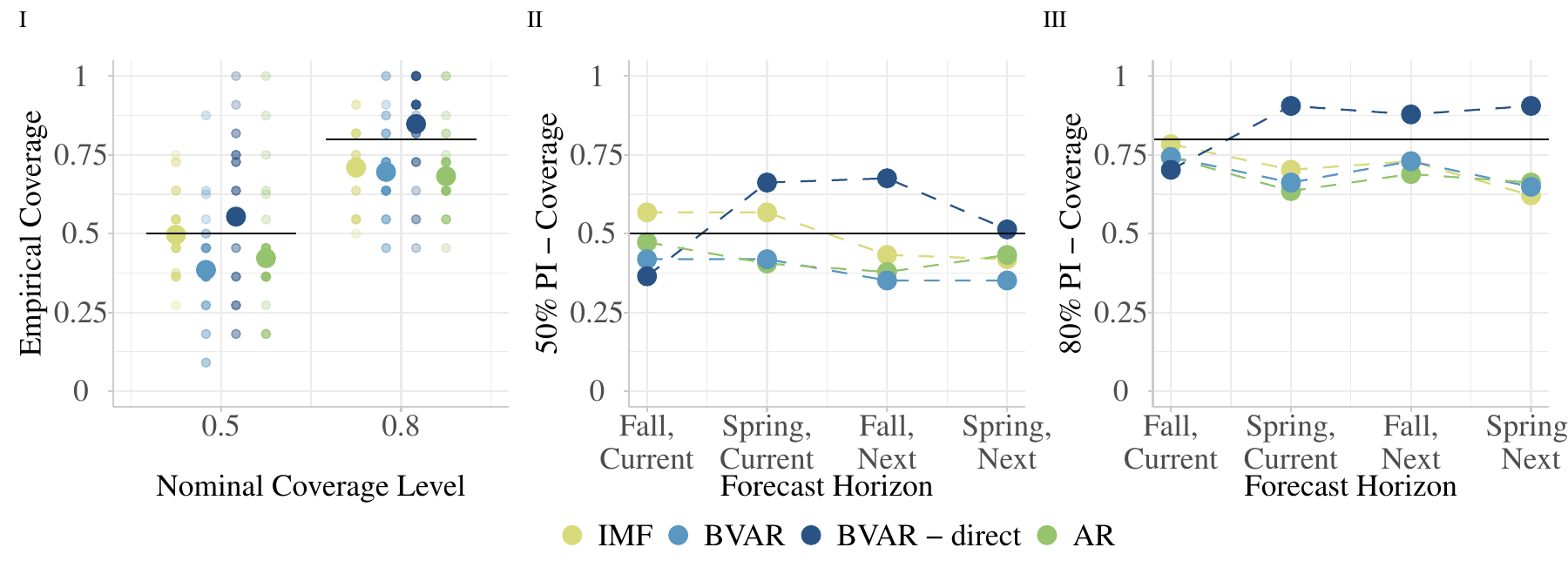}
		\caption{Inflation} 
	\end{subfigure}
	\vfill 
	
	\begin{subfigure}{\linewidth}
		\centering
		\includegraphics[width=1.01\linewidth]{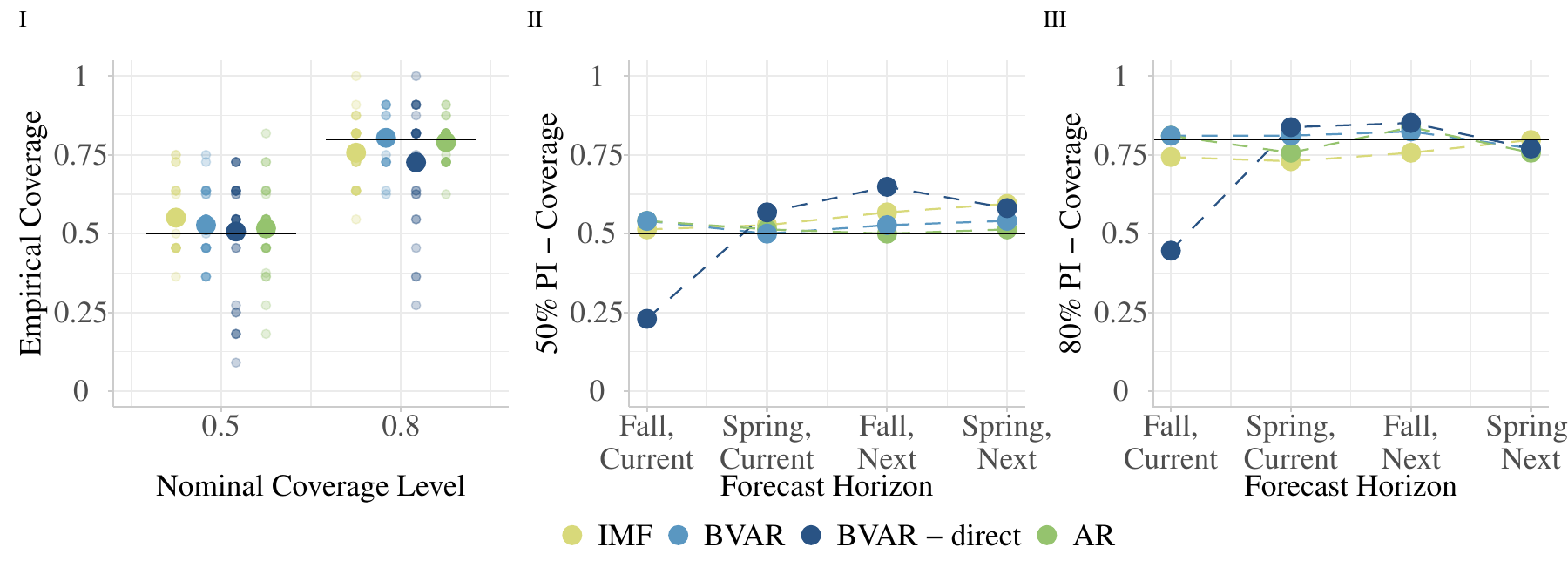}
		\caption{GDP growth} 
	\end{subfigure}
	\caption{Empirical prediction interval coverage levels, for (a) inflation and (b) GDP growth, evaluation period 2013-2023. In each panel, (I) shows the overall empirical coverage for all methods, with translucent smaller points in the background representing individual coverage for each country-horizon pair, indicating underlying variability in coverage levels. (II) and (III) show coverage levels by forecast horizon, for 50\% and 80\% nominal coverage, respectively. The horizontal lines indicate the respective nominal coverage level.} \label{fig:ho_coverage_absolute_rollingwindow}
\end{figure}

Here we use the 2013-2023 hold-out data for evaluating the model configuration ($R = 11$, absolute error method) chosen as described above.\footnote{As the inflation data series that we use for the benchmark models stops for Japan in 2021 and we thus lack benchmark predictions from 2021 onward, Japan is excluded from scoring in the years 2021 to 2023.}  As noted in Section \ref{sec:data}, we use the first fall release for defining the truth value against which we evaluate the predictions. Figure \ref{fig:ho_wis_absolute_rollingwindow} shows interval scores, whereas Figure \ref{fig:ho_coverage_absolute_rollingwindow} shows empirical coverage of the intervals.

For inflation, IMF-based forecasts tend to perform best at the shortest horizon (``Fall, Current"), whereas scores are otherwise similar for all methods. The decomposition into the score components is also similar, although skewed more towards dispersion for the benchmarks and towards underprediction for IMF-based forecasts, especially at next-year horizons. When considering interval scores separately at the two levels, see Table \ref{tab:baseis5080} in the supplement, the IMF-based forecasts are favored at the $50\%$ level, while the benchmarks tend to attain lower score values for the $80\%$ level, especially at the two longer horizons. Empirical coverage values overall tend to stay close to nominal levels, for all methods.

For GDP growth, the IMF-based forecasts tend to receive better interval scores than competing methods, across all horizons. This ordering is also largely upheld when considering interval scores separately at the two levels, see Table \ref{tab:baseis5080} in the supplement.
The score decomposition is similar for next-year forecasts across methods, with overprediction accounting for the largest portion, followed by dispersion. Recall that the over- and underprediction components scale linearly with the distance of forecast interval and realized value. Thereby, years such as 2020 with pronounced negative growth rates, which generally were not anticipated a year or more in advance, enter particularly heavily into the overprediction component. Concerning calibration, all methods attain close-to-nominal empirical coverage levels. The IMF-based forecasts overall show little variability in empirical coverage across horizons and countries, but are slightly less well calibrated at the $50\%$ level than competing methods.

We also consider \cite{DieboldMariano1995} for the null hypothesis of equal predictive accuracy of the IMF-based method and each benchmark method. In order to limit the number of tests conducted, we average the interval scores across countries at each date (for a given variable, forecast horizon and benchmark model). Pooling information in this way can further provide greater statistical power in empirical applications such as ours \citep{QuEtAl2023}. This yields 24 tests in total (two variables, four horizons, and the AR, BVAR and BVAR-direct benchmark models). We consider two-sided tests, a significance level of 95\% and use the \textsf{R} package \textsf{sandwich} and its function \textsf{NeweyWest} \citep{Zeileis2004,ZeileisEtAl2020} for the autocorrelation-robust variance estimator that enters the test statistic. This test statistic corresponds to $J_{n,T}^{\mathrm{DM}}$ in \cite{QuEtAl2023} and is a special case of the test statistic $S_{nT}^3$ in \cite{Akgun2024}. Among the 24 test setups, the null of equal predictive accuracy is rejected in favor of the IMF forecasts on four occasions (all referring to GDP growth at long horizons), whereas there are no rejections in favor of a benchmark model. We further consider $\tilde{S}_{nT}^3$ from \cite{Akgun2024}, which does not require $T$ to be large but also cannot account for serial correlation in the loss differentials of the test statistic. $\tilde{S}_{nT}^3$ relies on the critical values of the Student's t-distribution which leads to a more conservative result, with only one rejection in favor of the IMF forecasts.\footnote{The p-values change from 0.7\%, 2.5\%, 2.8\% and 0.6\% to  5.4\%, 19.4\%, 17.5\%, 1.0\%, respectively.} These results partly reflect the short time series dimension of our evaluation sample and the associated low power of the tests.

All in all, we conclude that IMF-based forecasts attain similar or sometimes slightly better performance than competing methods, for both targets and main evaluation metrics used.

\begin{figure}
\centering
  \includegraphics[width=0.85\linewidth]{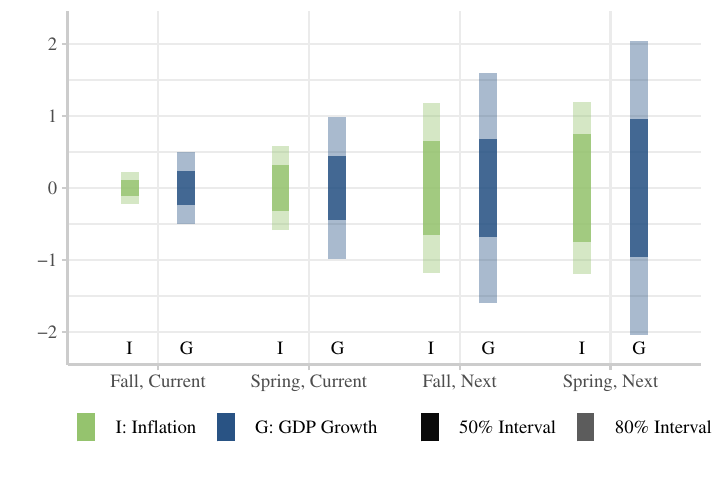}
\caption{Average length of prediction intervals for IMF-based probabilistic forecasts, in the years 2013-2023.} \label{fig:horizon_unc}
\end{figure}

Figure \ref{fig:horizon_unc} shows the average length of the prediction intervals of IMF-based probabilistic forecasts, separately for each combination of target variable and forecast horizon. For both inflation and GDP growth, lengths shrink considerably between next-year and same-year horizons, as well as between the two same-year horizons. The reduction in size is less considerable between the two next-year horizons, especially for inflation, suggesting that the average level of confidence stays mostly similar from one-and-a-half years to one year out from the target. This result is in accordance with \cite{qu2019any}, who do not find consistent significant evidence that the WEO's average forecast accuracy improves across these horizons.

\subsection{Further Analyses}
\label{sec:further_analyses}

Here we summarize further analyses that investigate the robustness of our empirical results (Sections \ref{sec:oecdtruth}--\ref{sec:addcountries}) and illustrate the properties of our proposed approach (Section \ref{sec:casestudies}). 

\subsubsection{OECD Truth Values}
\label{sec:oecdtruth}

The results in Figures \ref{fig:ho_wis_absolute_rollingwindow} and \ref{fig:ho_coverage_absolute_rollingwindow} are based on annual truth values provided by the IMF. By contrast, the benchmark models described in Section \ref{sec:bench} are fitted to higher-frequency (quarterly) data that are not provided by the IMF, and were thus chosen from another source (OECD). While we made our best effort to choose compatible series, full consistency between the two sources seems unrealistic given the complexity of international macroeconomic aggregates. As a consequence, our use of IMF data for evaluation may pose an unfair disadvantage to the benchmark models that were fitted on OECD data. To examine this possibility, we consider an alternative set of annual truth data that we construct from quarterly OECD data. We use these truth data for both the calculation of forecast errors and the subsequent evaluation. Figure \ref{fig:oecdvsimf} in the supplement studies how this change affects interval scores. The figure focuses on the comparison between the IMF-based forecasts and BVAR-direct as an illustrative benchmark method. In a majority of setups, the different choice of truth value does not change the model ranking between the IMF and BVAR-direct. However, for the shortest horizon (``Fall, Current"), the model ranking is typically reversed in favor of the BVAR-direct method. These results are confirmed by Figures \ref{fig:ho_oecd_wis_absolute_rollingwindow} and \ref{fig:ho_oecd_coverage_rollingwindow} in the supplement. Thus, while the IMF-based forecasts remain competitive at the other forecast horizons, their good performance at the shortest horizon vanishes when using OECD rather than IMF truth values. While somewhat unsatisfactory, this finding is plausible from a statistical perspective: The shortest forecast horizon is characterized by fairly high predictability since relevant components of the predictand (e.g., GDP growth in 2022) are known when the forecast is issued (e.g., September 2022). In this low-noise setting, the details of the definition of the truth value matter. Hence the IMF-based model outperforms the benchmarks when evaluated against IMF data, whereas the benchmarks prevail when evaluated against OECD data. This effect is not present at longer forecast horizons, where predictability is much lower, so that differences in the definition of the truth value are dominated by other sources of forecast error. 

\subsubsection{Additional Intervals and Measures of Forecast Performance}

In the main part of the paper, our analysis of forecast performance has focused on prediction intervals at the 50\% and 80\% level, i.e., $\tau \in \{0.5, 0.8\}$, which we consider particularly relevant in practical applications. Here, we consider an extension where we generate prediction intervals at equally spaced levels $\tau \in \{0.1, 0.2, ..., 0.9\}$, and additionally report more evaluation measures, to investigate robustness of results with respect to these extensions. 

In Tables \ref{tab:extcis_currentyr} and \ref{tab:extcis_nextyr} in the supplement, we report the evaluation measures for these forecast intervals, including various types of interval scores as well as the Continuous Ranked Probability Score (CRPS; \citealt{MathesonWinkler1976}) that each consider the entire predictive distribution. In addition, some individual interval scores at representative levels $\tau \in \{0.3, 0.5, 0.8, 0.9\}$ are shown in Table \ref{tab:extcis_isscores} in the supplement. These measures yield the same qualitative conclusions as our previous analysis. 

Additionally, Figure \ref{fig:extcis_qucvg} in the supplement presents coverage statistics for the corresponding forecast quantiles, where a forecast interval with $\alpha = 1-\tau$ induces the forecast quantiles $\hat{q}_{\alpha/2}$ and $\hat{q}_{1 - \alpha/2}$.

As shown there, the proposed IMF-based method attains close-to-nominal calibration in most instances, whereas some of the benchmark models exhibit considerable deviations from nominal coverage, especially for GDP growth. Nevertheless, for inflation and quantile levels larger than 0.85, some benchmarks, e.g., BVAR-direct, attain better calibration than the IMF-based forecasts.

\subsubsection{Additional Benchmark Models}\label{sec:add_bench}

In addition to the AR(1) and BVAR models described earlier, Tables \ref{tab:extcis_currentyr} and \ref{tab:extcis_nextyr} in the supplement present results for additional benchmark models. Results are shown for both the extended set of intervals introduced in the previous section and the base set of intervals from the main analysis. Conceptually, these models are variations and extensions of the main benchmarks, to investigate robustness of results with respect to concrete benchmark specifications. 
These include an AR($p$) model, where the lag length parameter $p$ is selected recursively using the \cite{Schwarz1978} information criterion; the `BVAR-CISS' model including a summary indicator of financial stress \citep{HolloEtAl2012}; the `BVAR-Mix' model that uses stochastic volatility prior to 2019 and constant volatility afterwards (see Section \ref{sec:bench}); and an ensemble using equally weighted quantile-wise averages of the IMF, AR and BVAR forecasts. The ensemble is a natural benchmark given extensive empirical evidence on the good performance of forecast combinations in various empirical setups \citep{WangEtAl2023}. In addition to benchmarks trained on quarterly OECD data, we fit two autoregressive specifications on annual outcome data provided by the IMF, and thereby tailor models directly to the annual growth rates to be predicted. While `AR-annual' only has access to the latest annual truth value, `ARX-annual' compensates for this informational disadvantage by incorporating the most recent IMF point forecasts as regressors. The latter is thus based on the same information set as our proposed approach. Section \ref{sec:details_bench} in the supplement provides more detailed descriptions of the additional benchmarks. 

We find that previous results are in general robust to concrete specifications of the benchmark models. `AR-annual' exhibits high scores on current-year forecasts, where its informational disadvantage is particularly high, but performs comparably to other models on next-year forecasts. In terms of the CRPS, some alternative specifications outperform the respective counterparts from the main analysis on a subset of target-horizon combinations. In particular, the AR($p$) specification outperforms the AR(1) variant for inflation at the next-year horizons. Overall, however, any improvements of alternative specifications over the main benchmarks tend to be small and inconsistent across setups.

Compared to our proposed IMF-based approach, the BVAR-Mix model performs slightly better for inflation at both next-year horizons, but performs similar or worse otherwise. This model additionally requires a manual choice of regime switch (necessitated by Covid-19 outliers) which can be viewed as a conceptual drawback as compared to the other models we consider. The ARX-annual method performs worse than the IMF-based approach in most setups. Given that both methods are based on the same information set, this implies that the modeling assumptions of the IMF-based approach are more suitable from a forecasting perspective. 

Among all benchmarks, the ensemble tends to perform best overall, and similarly well as our proposed IMF-based method.

\subsubsection{Additional Countries}
\label{sec:addcountries}
An attractive feature of our methodology is its flexibility to be readily extended to generate prediction intervals for additional countries. This is particularly appealing in conjunction with the IMF WEO point forecasts, which have nearly worldwide coverage.

Thus, while our focus remains on the G7 economies, we extend a version of our analysis to ten additional countries in order to investigate the robustness of our approach. The set of benchmark models here is reduced to those that are fit directly on annual IMF truth data. We choose each of the five largest Emerging Market Economies and Advanced Economies (besides the G7), with data on country classification and economy size obtained from the IMF \citep{IMFWEOeconomies}. These criteria yield the following countries:  China, India, Brazil, Russia and Mexico (Emerging Market Economies) and South Korea, Australia, Spain, the Netherlands and Switzerland (Advanced Economies).
We thereby keep focus on a comparable number of countries to the originally considered G7, while still retaining some generality across different economy types.  

As shown in Figures \ref{fig:cvg_extctry_pcpi_pch} and \ref{fig:cvg_extctry_ngdp_rpch} in the supplement, our proposed approach attains close-to-nominal coverage for the extended set of countries as well. While aggregate coverage levels are in some cases slightly better for the AR-annual and ARX-annual models, the IMF-based forecasts exhibit lower levels of variability in empirical interval coverage across countries. Variability across horizons is similarly small for all methods. In terms of different variants of the interval score, see Tables \ref{tab:extctry_pcpi_pch} and \ref{tab:extctry_ngdp_rpch} in the supplement, our proposed method consistently outperforms both benchmarks, and in particular the ARX-annual method that is based on the same information set. 

\subsubsection{Forecasts During the Covid-19 Period}
\label{sec:casestudies}

Tables \ref{tab:sens_cov19gdp} and \ref{tab:sens_cov19cpi} in the supplement present our main performance measures for the period 2020--2023, that is, the Covid-19 pandemic period and its immediate aftermath. We present results for this sub-period in order to investigate whether or not our results are driven by a small number of extreme observations. That said, this analysis should be viewed as illustrative since conditioning on extreme events is subject to the ``forecaster's dilemma'' critique \citep{lerch2017} and is thus not helpful in terms of formal forecast evaluation.

For GDP growth, the IMF-based forecasts continue to outperform the benchmarks in terms of different variants of the interval score, for three out of four horizons. For inflation, the IMF-based forecasts perform worse than the benchmarks in all but the shortest horizon. In conjunction with the interval score results for the entire hold-out sample (2013--2023, see Figure \ref{fig:ho_wis_absolute}, and additionally Tables \ref{tab:extcis_currentyr} and \ref{tab:extcis_nextyr} in the supplement), this finding indicates that our proposed IMF-based method generally outperforms the benchmarks for the 2013--2019 period, whereas the ranking for the Covid-19 period is mixed as described. Unsurprisingly, all methods attain less--than-nominal quantile coverage rates during the Covid-19 period (on average across years and countries). That said, the coverage rates of the IMF-based method are clearly lower than for the benchmark methods. 

Furthermore, the Covid-19 period is interesting to showcase how our proposed method's prediction intervals are updated over time. Compared to some of the benchmark models, our method's updating process is rather insensitive to shocks, which is due to two main features. First, the use of quantiles, which clearly limits the impact of extreme observations. Second, the use of annual (as opposed to quarterly) data on forecast errors. Figures \ref{fig:cov19ints_hr0gdp}, \ref{fig:cov19ints_hr1gdp}, and \ref{fig:cov19ints_hr05cpi} in the supplement illustrate this point by showing prediction intervals before and after years with large forecast errors (2020 for GDP growth and 2022 for inflation).

From an ex-post perspective, low responsiveness to Covid-19-related shocks turned out advantageous in the case of GDP growth, where the extreme forecast error in 2020 was followed by moderate forecast errors afterwards. For inflation, where IMF-based intervals were already quite narrow compared to the benchmark models, the low immediate reactiveness of the quantile-based method was a disadvantage, especially if the increase in forecast errors after 2022 represents a continued structural change beyond the end of our study period, which ends in 2023. Within the scope of our prospective study, see Section \ref{sec:shiny}, we will be able to monitor the further performance of IMF-based forecast intervals for inflation.

Apart from statistical forecasting performance, our method's use of quantiles and annual data clearly contributes to its simplicity and transparency. By contrast, the forecasts and updating mechanisms of statistical and machine learning methods tend to be far more complicated and may depend on hyperparameters in non-obvious ways. 

\subsection{Outlook: Prospective Real-Time Forecasts and Evaluation}
\label{sec:shiny}
In addition to the analysis presented so far, we plan to employ the same method in real time and make the resulting probabilistic forecasts publicly available. This is motivated by a lack of easily available probabilistic forecasts for some G7 countries, as discussed in the introduction. 

Initial biannual releases of the forecasts starting October 2023 are available in a public GitHub repository \citep{macropigit}, ensuring transparency and accountability through time stamps. The latter repository also links to the website \url{https://probability-forecasting.shinyapps.io/macropi/} on which we provide concise visualizations of all forecasts and prediction intervals. Moreover, we have deposited a preregistration protocol \citep{osf_prereg} that covers forecasts from October 2024 onwards. In this protocol, we make a commitment regarding how we will evaluate forecasts in 2029, after collecting prospective forecast intervals for five years.

\section{Conclusion}\label{sec:conc}

This paper presents a method for predicting the distribution of output growth and inflation in the G7 economies. Our method aims to be as simple as possible, subject to being ``reasonably competitive'', in the sense that we see no straightforward way to improving empirical forecasting performance. In particular, assessing the impact of our method's tuning parameters (the number of past forecast errors considered, as well as the use of absolute versus raw forecast errors) is easier than in time series models that achieve comparable forecast performance. Among the benchmark methods we consider, an ensemble of our proposed approach with two time series models performs similarly well in terms of overall forecast accuracy. At the same time, the ensemble is harder to replicate and communicate in that obtaining and explaining its forecasts requires input from all three component forecasts. From the perspective of forecast performance, potential limitations of our quantile-based method include its limited consideration of extreme events and a comparably slow adaptation to potential structural change in the variability of forecast errors. This behavior, however, has also led to some robustness of results e.g. during the Covid-19 period but will be more closely examined in our prospective study. 

While we use point forecasts from the IMF, our suggested procedure can in principle be employed for point forecasts from any source, including macroeconomic model-based forecasts, judgmental forecasts (issued, e.g., by central banks and surveys), forecasts based on statistical or machine learning models, possibly using a large number of input variables. This versatility is useful in practice, where point forecasts are often readily available, be it via existing data sources (as for macroeconomic survey forecasts) or software packages (as for model based forecasts). As an important and often critical practical requirement, an informative history of out-of-sample forecast errors must be available in order to estimate the distribution of forecast errors. From an economic forecasting perspective, our empirical results on the IMF forecasts display the common finding that point forecasts by economic institutions are not easy to beat by means of purely statistical models \cite[see e.g.][]{FaustWright2013}. 

\bibliographystyle{ecta}

\newpage
\appendix
\section*{Supplement}
\section{Uncertainty Quantification in Central Bank Reports} \label{sec:cb_reports_detailed}

\subsection*{Protocol for Extracting Information about Uncertainty Quantification}
We are searching for concretely quantified statements about the uncertainty surrounding macro-economic forecasts in the G7's central banks' ``flagship'' reports for communicating their growth and inflation projections. For this, we applied the following protocol. \\ 
We note that this protocol does not necessarily provide a comprehensive view about central banks' handling of prediction uncertainty and that each central bank might issue explicitly quantified statements about the uncertainty surrounding their macroeconomic forecasts elsewhere. The protocol thus only provides a snapshot view of whether and how central banks communicate uncertainty in their ``flagship'' reports for macroeconomic projections.
\begin{enumerate}
\item \textbf{Selection of Document.} \\ For each country, we searched (via the Google search engine) for ``[the respective central bank]'' + ``Macroeconomic Projections''. Subsequently, we selected the first pdf document that stemmed from that central bank and contained forecasts for both GDP growth and inflation. Often, this document would carry the name ``Macroeconomic Projections'', alternatively usually "Monetary Policy Report" or ``Economic Outlook".

\item \textbf{Information extraction given document} 
\begin{enumerate}
\item \textbf{Information around forecasts}\\ Via a combination of manual and automated search, we navigated to the place in the document that contained macroeconomic forecasts for GDP growth and inflation. These would typically be communicated in a graphical and/or table format. For each graph or table, we checked whether any information additional to a forecast of central tendency was supplied. We also checked the directly surrounding text. 
\item \textbf{Searching for key terms}\\ To avoid missing any additional information concerning uncertainty around the communicated forecasts, we additionally searched the document for the following key terms:
\begin{itemize}
\item uncertainty 
\item confidence (bands)
\item quantile
\item mean
\item median
\item probabilistic
\item range 
\item interval
\end{itemize}
Given a match, we determined from context whether the given statement contained any quantification of prediction uncertainty for the macroeconomic forecasts in the document.
\end{enumerate}
\end{enumerate}
\begin{landscape}
\begin{table}
\begin{center}
\begin{tabular}{l l p{4.25cm} l c p{3.5cm}}
 \textbf{Country} &  \textbf{Name of Central Bank} &\textbf{Name of Document}  & \textbf{Edition} & \textbf{UQ present} & \textbf{Form of UQ}\\[0.5em]
 \hline \\[-1em]
 Canada & Bank of Canada & Monetary Policy Report & January 2025 & no & - \\  
 France & Banque de France & Macroeconomic Projections France & December 2024 &no &- \\ 
 Germany & Deutsche Bundesbank & Monthly Report & December 2024\footnotemark[1]  & no &- \\ 
 Italy & Banca d'Italia & Macroeconomic Projections for the Italian Economy & April 2025 & yes & fan chart \\  
 Japan & Bank of Japan & Outlook for Economic Activity and Prices & January 2025 & somewhat & truncated range of board members' votes\footnotemark[2] \\ 
 United Kingdom \hspace*{5mm}& Bank of England & Monetary Policy Report & February 2025
& yes &fan chart \\ 
 United States & Federal Reserve & Monetary Policy Report & February 2025 & yes &multiple, including fan chart    
 \end{tabular}
\caption{State of uncertainty quantification for macroeconomic forecasts that are issued by the central banks of the G7 countries. For each central bank, we consulted the latest available ``flagship'' report containing forecasts for inflation and real GDP growth, often titled ``Macroeconomic Projections'', ``Economic Outlook'' or ``Monetary Policy Report'' (depending on availability) and checked whether that document's centrally communicated forecasts are accompanied by an explicit quantification of uncertainty. All reports were retrieved from the respective bank's website on April 14, 2025.\\ 
$^1$\footnotesize{Not all monthly reports contain macroeconomic forecasts - the most recent publication containing them was thus chosen.}\\ 
$^2$\footnotesize{These can't be interpreted as quantiles of the forecast distribution, as every board member is presumably forecasting a central functional (mean, median, etc.) of the forecast distribution.} }
\label{tab:cbreports_detailed}
\end{center}
\end{table} 
\end{landscape}

\clearpage

\section{Details on Empirical Results} \label{sec:appendix_empirical}

\subsection{Data for Benchmark Models}\label{sec:bmdata}

CPI: quarterly index data downloaded from \url{https://stats.oecd.org} on March 18, 2024; OECD identifier: \verb|CPALTT01|. We compute inflation rates as logarithmic growth rates of the CPI levels provided by the raw data.\\
GPD: quarterly growth rates of real GDP downloaded from \url{https://data-explorer.oecd.org/} on March 21; OECD identifier: Table \verb|0102|.\\
CISS: daily values of the CISS index downloaded from \url{https://data.ecb.europa.eu/} on March 18, 2024; series identifier: \verb|CISS.D.[country code].Z0Z.4F.EC.SS_CIN.IDX|, where country code is a country code (DE = Germany, FR = France, GB = Great Britain, IT = Italy, US = United States of America). We compute quarterly levels of the CISS index by averaging all daily levels within the quarter.\\
For evaluating forecasts of CPI and GDP, we obtain annual observations as a weighted sum of seven quarterly growth rates, as detailed e.g. in \cite{PattonTimmermann2011} or Section A.2 of the online supplement for \cite{KruegerPlett2023}. We also use this representation to compute BVAR forecast distributions of the annual predictands, based on the quarterly data to which the BVAR models are fitted. 

\subsection{Individual Interval Score Results}
\begingroup
		\renewcommand{\arraystretch}{1.13}
\begin{table}[!h]
\centering
\caption{Individual interval scores for main analysis at the $50\%$ and $80\%$ level, evaluation period 2013--2023. These are the scores underlying the weighted sum in Figure \ref{fig:ho_wis_absolute}.}
\label{tab:baseis5080}
\begin{tabular}{p{0.8cm}lllll}
\toprule
\multicolumn{2}{c}{ } & \multicolumn{2}{c}{{Inflation}} & \multicolumn{2}{c}{{GDP Growth}} \\
\cmidrule(l{3pt}r{3pt}){3-4} \cmidrule(l{3pt}r{3pt}){5-6}
 &  & $\text{IS}_{50}^{1)}$ & $\text{IS}_{80}^{1)}$ & $\text{IS}_{50}^{1)}$ & $\text{IS}_{80}^{1)}$\\
\midrule
\parbox[t]{2mm}{\multirow{4}{*}{\rotatebox[origin=c]{90}{\parbox{2cm}{\centering Fall,\\Current}}}} & IMF & \textbf{0.46} & \textbf{0.71} & \textbf{1.25} & 2.27\\
 & AR & 0.83 & 1.46 & 1.47 & 2.28\\
 & BVAR & 0.86 & 1.52 & 1.4 & \textbf{2.21}\\
 & Direct$^{2)}$: BVAR & 0.86 & 1.54 & 1.43 & 2.49\\
\addlinespace
\parbox[t]{2mm}{\multirow{4}{*}{\rotatebox[origin=c]{90}{\parbox{2cm}{\centering Spring,\\Current}}}} & IMF & \textbf{1.8} & 3.24 & \textbf{2.21} & \textbf{3.79}\\
 & AR & 1.96 & \textbf{3.2} & 3.49 & 6.13\\
 & BVAR & 1.96 & 3.28 & 3.1 & 5.26\\
 & Direct$^{2)}$: BVAR & 2.06 & 3.35 & 3.03 & 5.16\\
\addlinespace
\parbox[t]{2mm}{\multirow{4}{*}{\rotatebox[origin=c]{90}{\parbox{2cm}{\centering Fall,\\Next}}}} & IMF & \textbf{3.86} & 6.84 & \textbf{5.13} & \textbf{9.75}\\
 & AR & 4.57 & 6.65 & 6.35 & 11.46\\
 & BVAR & 4.35 & \textbf{6.28} & 5.88 & 11.07\\
 & Direct$^{2)}$: BVAR & 4.39 & 6.76 & 5.9 & 11.04\\
\addlinespace
\parbox[t]{2mm}{\multirow{4}{*}{\rotatebox[origin=c]{90}{\parbox{2cm}{\centering Spring,\\Next}}}} & IMF & \textbf{4.74} & 8.42 & \textbf{5.52} & \textbf{10.28}\\
 & AR & 5.95 & \textbf{7.96} & 7.51 & 12.97\\
 & BVAR & 5.76 & 8.04 & 7.26 & 13.27\\
 & Direct$^{2)}$: BVAR & 5.7 & 8.41 & 7.14 & 12.82\\
\bottomrule 
\multicolumn{6}{l}{\parbox{0.62\linewidth}{\scriptsize
	${1)}$ Different Versions of the Interval Score. $\text{IS}_{50}$ and $\text{IS}_{80}$ represent the interval score for the $50\%$ and $80\%$ intervals, respectively.}}\\
\multicolumn{6}{l}{\parbox{0.62\linewidth}{\scriptsize
	${2)}$ Forecast quantiles are taken directly from the parametric forecast distribution of the respective model.}}
\end{tabular}
\end{table}
\endgroup

\subsection{Training Sample Performance}
\subsubsection{Illustrative Analysis}
\label{sec:dir_illustrative}
Given the forecast intervals obtained from the two error methods in the training period 2001--2012, we empirically illustrate some of the general points mentioned in Section \ref{sec:errormethod} in the following.

\begin{figure}
  \begin{subfigure}{0.46\linewidth}
  \centering
    \includegraphics[width=\linewidth]{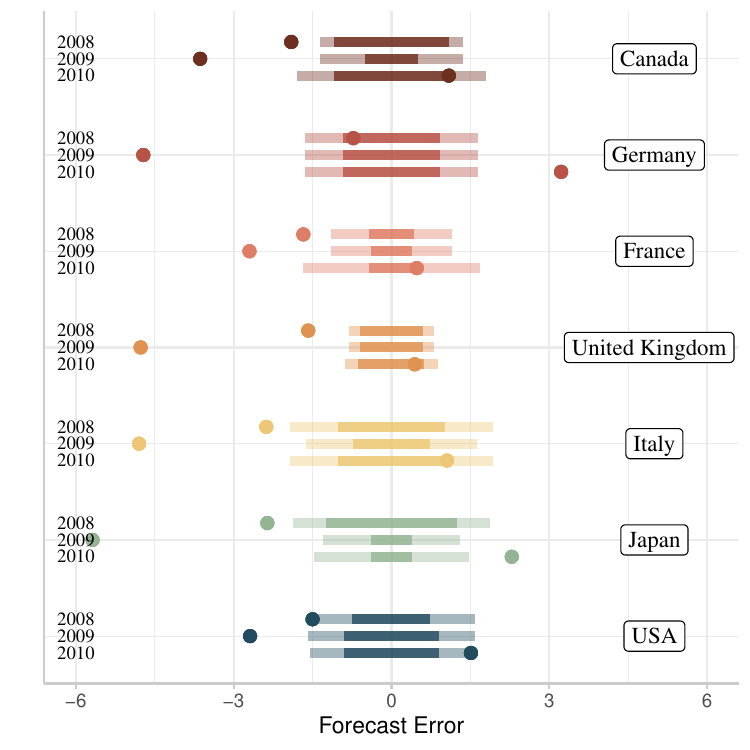}
    \caption{Absolute Errors}
  \end{subfigure}
  \hfill  
  \begin{subfigure}{0.46\linewidth}
  \centering
    \includegraphics[width=\linewidth]{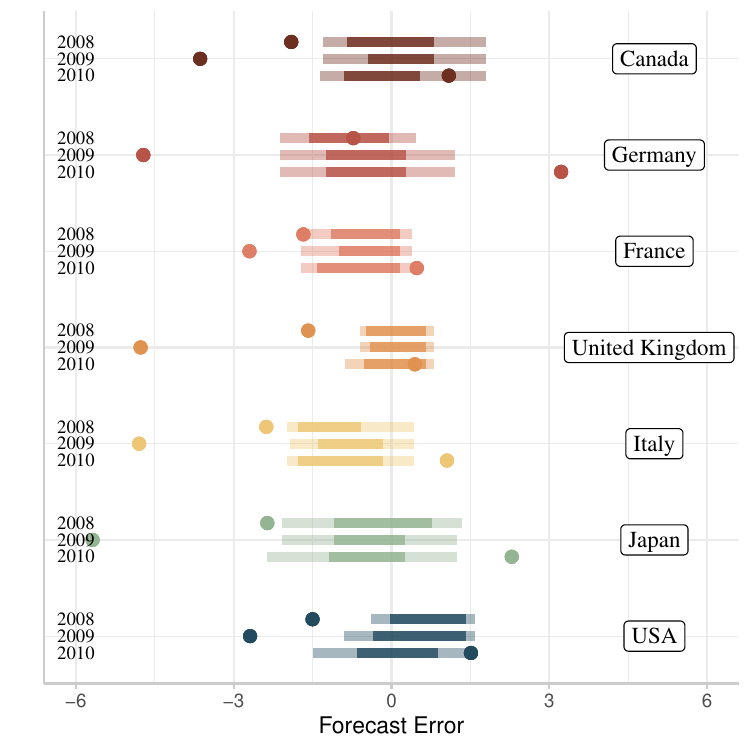}
    \caption{Directional Errors} 
  \end{subfigure}
\caption{Forecast error intervals for GDP growth in the G7 economies, for the "Fall, Next" horizon. Intervals are based on past IMF point forecast errors, for (a) the absolute error and (b) the directional error method. } 
\label{fig:dirvsabs_gdpfc}
\end{figure}

An attractive potential feature of the directional error method is that it is capable of shifting more probability mass to one side of the point forecast, thereby accounting for a potential skew in the forecast error distribution - for example, by better reflecting downward risk in GDP growth during periods of economic uncertainty. 
Figure \ref{fig:dirvsabs_gdpfc} illustrates the forecast error intervals for GDP growth in the G7 countries for the ``Fall, Next'' horizon during the years of the global financial crisis. In only 4 out of the 7 countries considered, the $80\%$ interval was in fact skewed to the negative side for both 2008 and 2009, which saw substantial negative forecast errors. Incorporating the forecast error from 2008, this shifted to 5 out of 7 countries for the target year 2010, which then however saw largely positive forecast errors. This point nicely illustrates that there is a potential mismatch in the duration and direction of economic uncertainty and the low frequency of annual data. Forecasts at the next-year horizons are particularly affected by this drawback, since there is a two-year gap between the last truth value that enters their information set and the completion of the target year.

\begin{figure}
\centering
  \begin{subfigure}{0.75\linewidth}
  \centering
    \includegraphics[width=\linewidth]{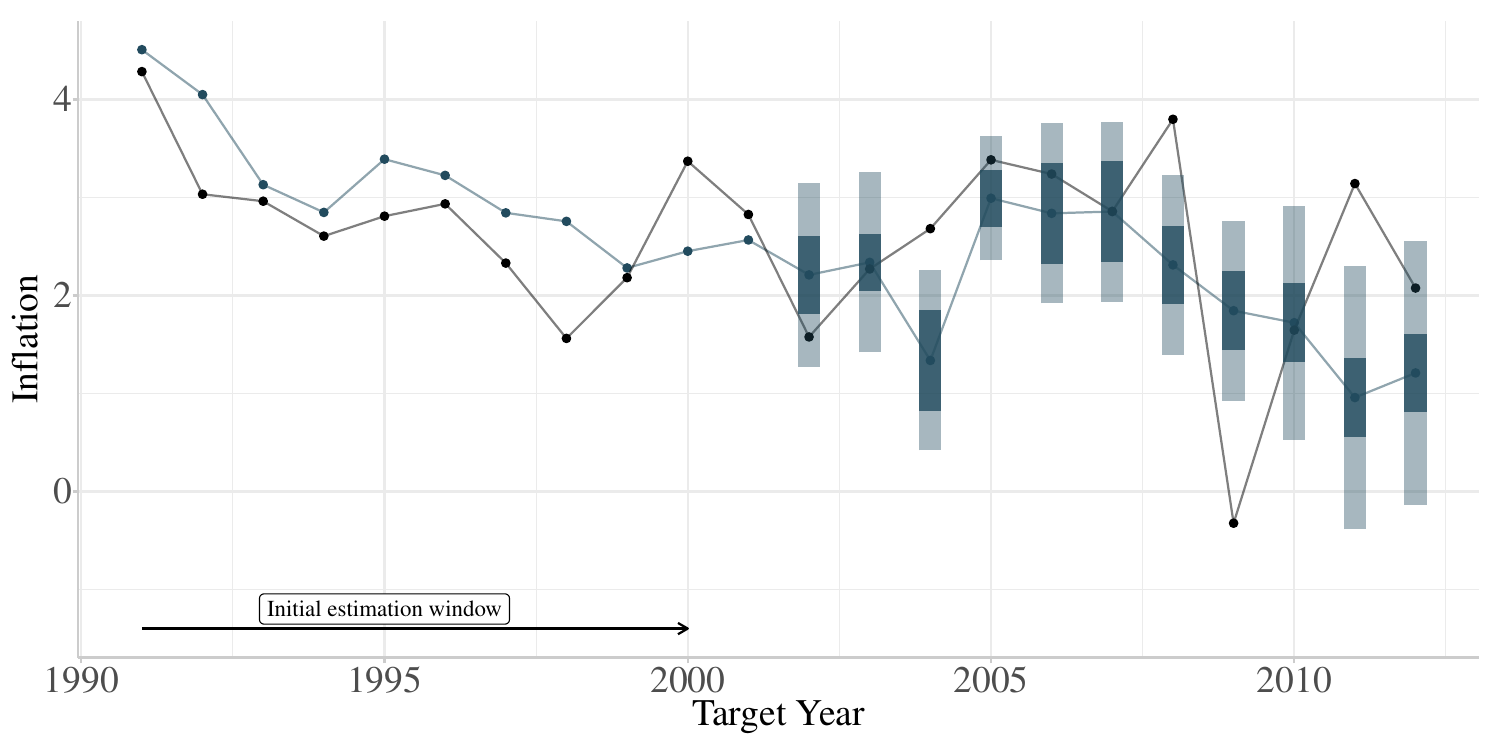}
    \caption{Absolute Errors} 
  \end{subfigure}
  \\
  \begin{subfigure}{0.75\linewidth}
  \centering
    \includegraphics[width=\linewidth]{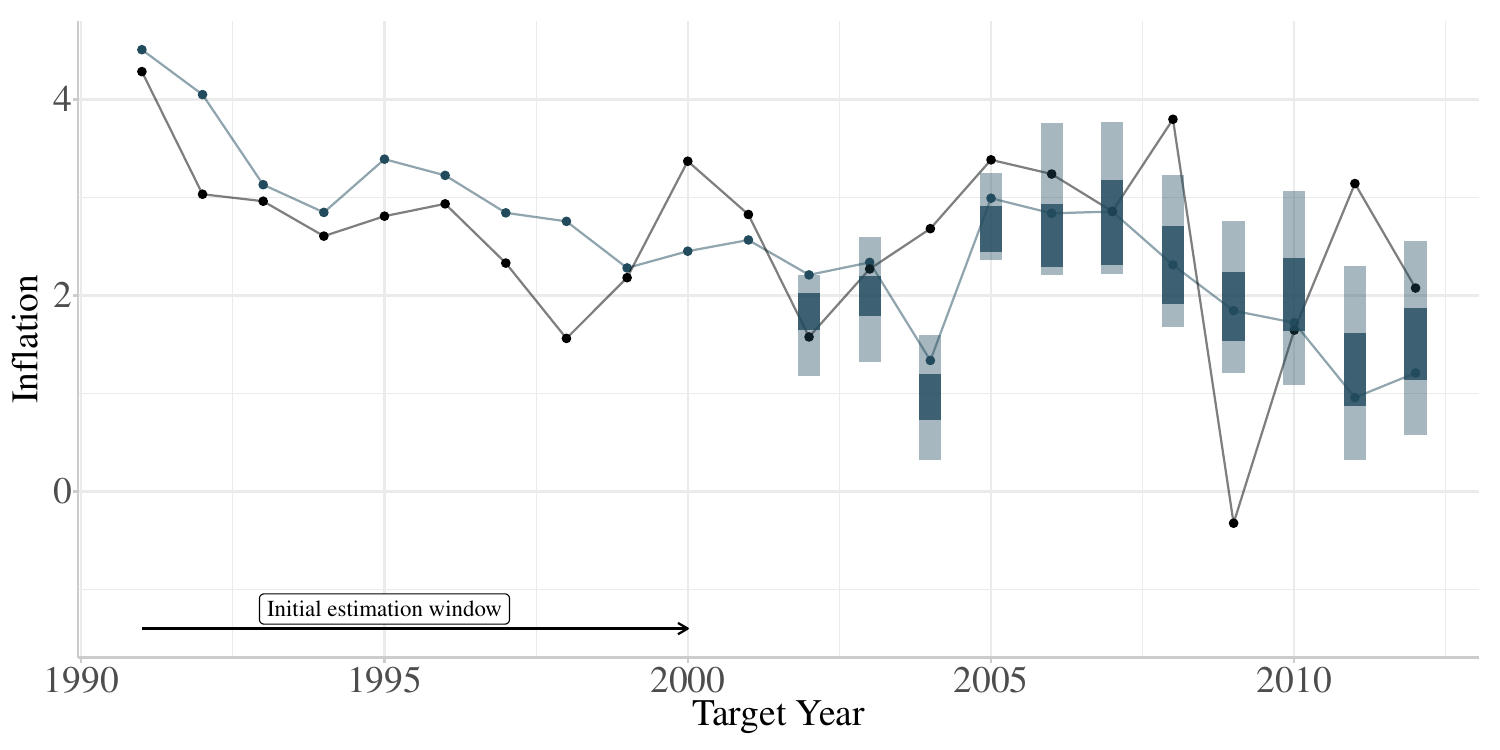}
    \caption{Directional Errors} 
  \end{subfigure}
\caption{Forecast error intervals for inflation in the USA, for the "Fall, Next" horizon. Intervals are based on past IMF point forecast errors, for (a) the absolute error and (b) the directional error method.} 
\label{fig:dirvsabs_USAcpi}
\end{figure}

Figure \ref{fig:dirvsabs_USAcpi} illustrates forecast intervals for the two methods for U.S. inflation in the ``Fall, Next Year'' horizon during the training period. During most of the initial estimation window, the IMF forecasts overestimate the actual inflation rate. The forecast intervals reflect this negative skew in the observed forecast errors and consequently shift most of the probability mass downward in the following years, which then however see an influx of positive forecast errors. We see this sort of behavior several times in our data, and argue that it is undesirable to only cover one direction of forecast uncertainty, especially when the corresponding estimate is based on a small number of observations and is therefore highly variable. This is further exacerbated by the fact that, at least in our setup, point forecasts are external and any potential past bias might be corrected for by the forecast issuer themselves, rendering any further correction for it questionable.

To sum up, while their potential to reflect a skew in the forecast error distribution might seem attractive, the  information contained in the signs of raw or directional forecast errors is variable due to a small number of observations. These observations are furthermore lagged (sometimes substantially), and potentially no longer representative of the current state of any directional forecast bias. This setup makes directional forecast errors an ineffective method to account for a possibly skewed distribution of the IMF's forecasts errors. 

\newpage
\subsubsection{Evaluation Metrics}
\begin{table}[h!]
\caption{Interval scores and interval coverage rates in the training sample (2001-2012), for the directional and the absolute error method.}
\label{tab:dirvsabs}
\begin{center}
\begin{tabular}{lcccccccccc}
\toprule
 & & &\multicolumn{2}{c}{$\text{IS}_{W,b}^{1)}$} & &\multicolumn{2}{c}{$\text{Cvg}_{50}^{2)}$} & &\multicolumn{2}{c}{$\text{Cvg}_{80}^{2)}$} \\[0.2em]
 \cmidrule(l{3pt}r{3pt}){4-5} \cmidrule(l{3pt}r{3pt}){7-8} \cmidrule(l{3pt}r{3pt}){10-11}
 &  && abs. & dir. & &abs. & dir. & &abs. & dir.\\[0.2em]
 \midrule
\multirow{4}{*}{\rotatebox[origin=c]{90}{GDP Growth}} & Fall, Current& & \textbf{0.23} & 0.24 & &\textbf{0.49} & 0.43 & &\textbf{0.76} & 0.65\\
 & Spring, Current && 0.41 & \textbf{0.41} & & 0.56 & \textbf{0.54} & &\textbf{0.76} & 0.67\\
 & Fall, Next && 0.91 & \textbf{0.88} & &\textbf{0.49} & 0.42 & &\textbf{0.73} & 0.70\\
 & Spring, Next && \textbf{1.14} & 1.15 & &\textbf{0.50} & 0.40 & &\textbf{0.64} & 0.55\\[1em]
\multirow{4}{*}{\rotatebox[origin=c]{90}{Inflation}} & Fall, Current && \textbf{0.12} & 0.12 & &\textbf{0.52} & 0.44 & &\textbf{0.76} & 0.64\\
 & Spring, Current && 0.26 & \textbf{0.25} && \textbf{0.43} & 0.39 && \textbf{0.75} & 0.65\\
& Fall, Next && \textbf{0.47} & 0.50 && \textbf{0.40} & 0.31 & &\textbf{0.67} & 0.54\\
 & Spring, Next && \textbf{0.52} & 0.55& & \textbf{0.42} & 0.38 & &\textbf{0.67} & 0.54\\
 \bottomrule
	\multicolumn{11}{l}{\parbox{0.75\linewidth}{\scriptsize
	${1)}$ A weighted sum of interval scores from quantile forecasts at base levels from main analysis $\tau \in \{0.5, 0.8\}$, with weights as given by \cite{bracher2020}.}}\\
	\multicolumn{11}{l}{\scriptsize
	${2)}$ Interval coverage at the $50\%$ and $80\%$ levels.}
\end{tabular}
\end{center}
\end{table}    

\begin{landscape}
\begin{figure}[!h]
\includegraphics[width=\linewidth]{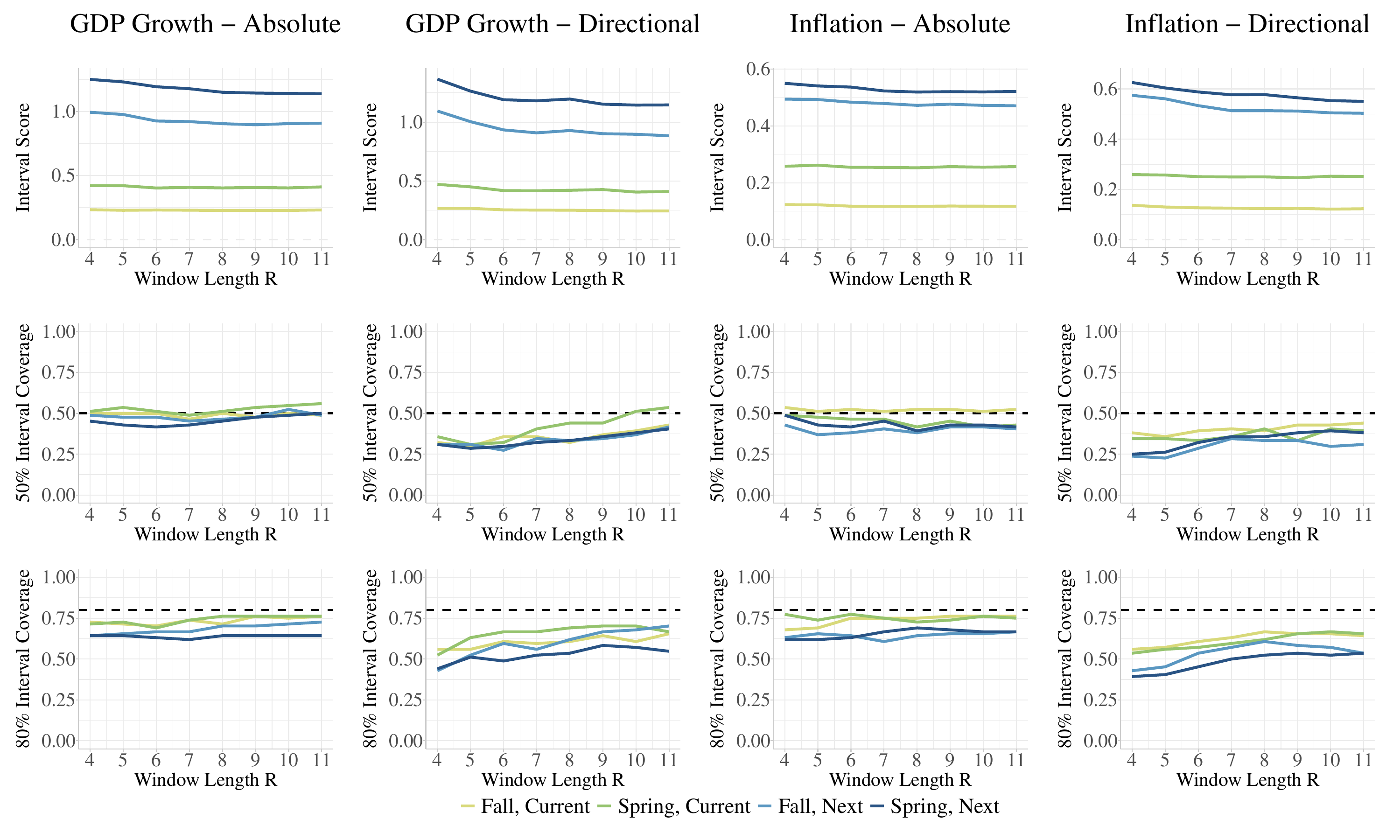}
\caption{Interval scores and empirical interval coverage rates in the training sample (2001-2012), for both targets and error extraction methods, and for various choices of the rolling window length $R$. Interval scores are a weighted sum of the interval scores at the $50\%$ and $80\%$ interval. The column header indicates the target and error extraction method. \label{fig:rwlength}}
\end{figure}
\end{landscape}
\clearpage

\subsection{Choice of Truth Source}
\begin{figure}[h]
\centering
  \includegraphics[width=0.85\linewidth]{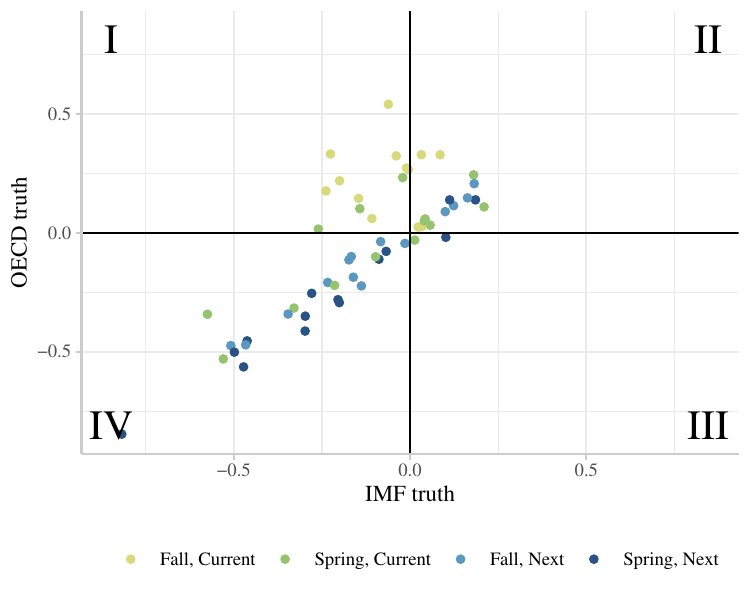}
\caption{Difference in interval scores (IS) of IMF-based and BVAR - direct forecasts $\text{IS}(\text{IMF}) - \text{IS}(\text{BVAR-direct})$, when using competing truth values as outcome data. Score differences from using IMF truth values are shown on the x axis, analogously for OECD truth values on the y-axis. Each data point corresponds to one combination of horizon, target variable, and country. Values in quadrants ``II'' and ``IV'' represent agreement between the two truth sources on which forecasts are better. Values in quadrant ``I'' represent instances where the IMF truth prefers the IMF forecasts, but the OECD truth prefers the BVAR forecasts. Values in quadrant ``IV'' represent instances where the IMF truth prefers the BVAR forecasts, but the OECD truth prefers the IMF forecasts.} \label{fig:oecdvsimf}
\end{figure}
\begin{figure}[h]
  \begin{subfigure}{0.475\linewidth}
  \centering
    \includegraphics[width=\linewidth]{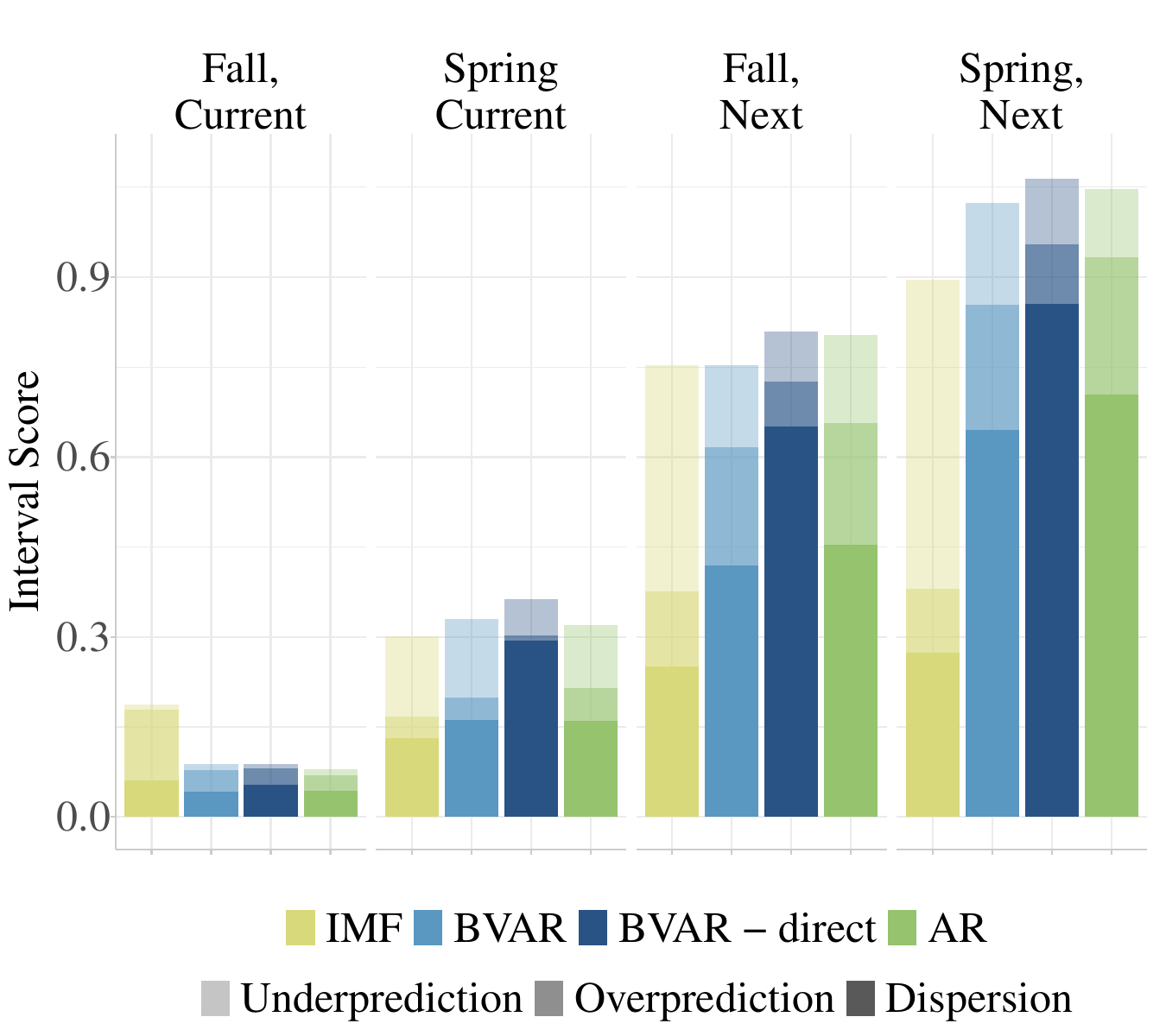}
    \caption{Inflation} \label{fig:oecd_wis_pcpi_pch_absolute_rollingwindow}
  \end{subfigure}
  \hfill  
  \begin{subfigure}{0.475\linewidth}
  \centering
    \includegraphics[width=\linewidth]{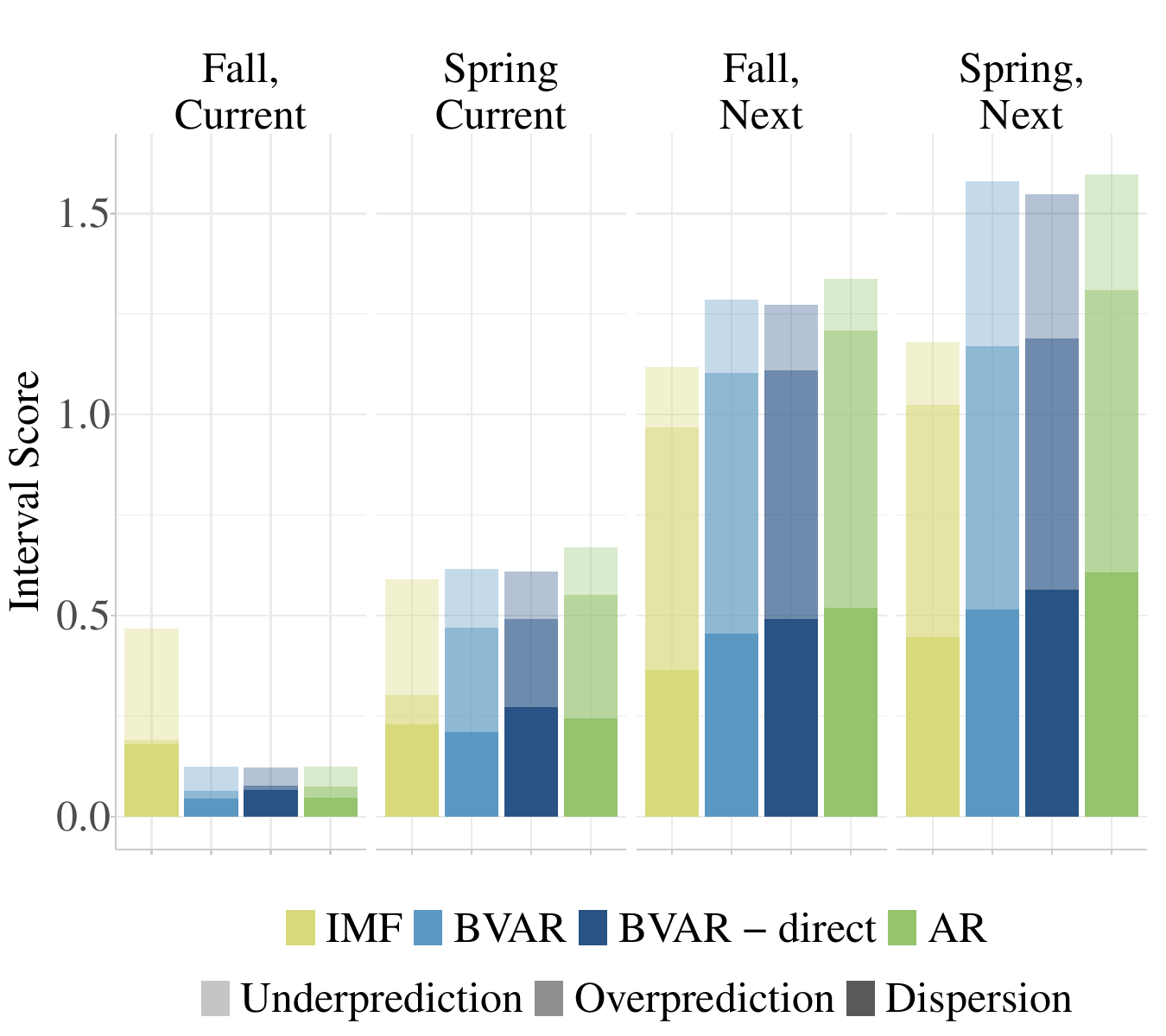}
    \caption{GDP growth} \label{fig:oecd_wis_ngdp_rpch_absolute_rollingwindow}
  \end{subfigure}
\caption{Same as Figure \ref{fig:ho_wis_absolute}, but using OECD truth values as the outcome data.} \label{fig:wis_absolute_oecd}
\label{fig:ho_oecd_wis_absolute_rollingwindow}
\end{figure}

\begin{figure}[h]
  \begin{subfigure}{\linewidth}
    \centering
    \includegraphics[width=1.01\linewidth]{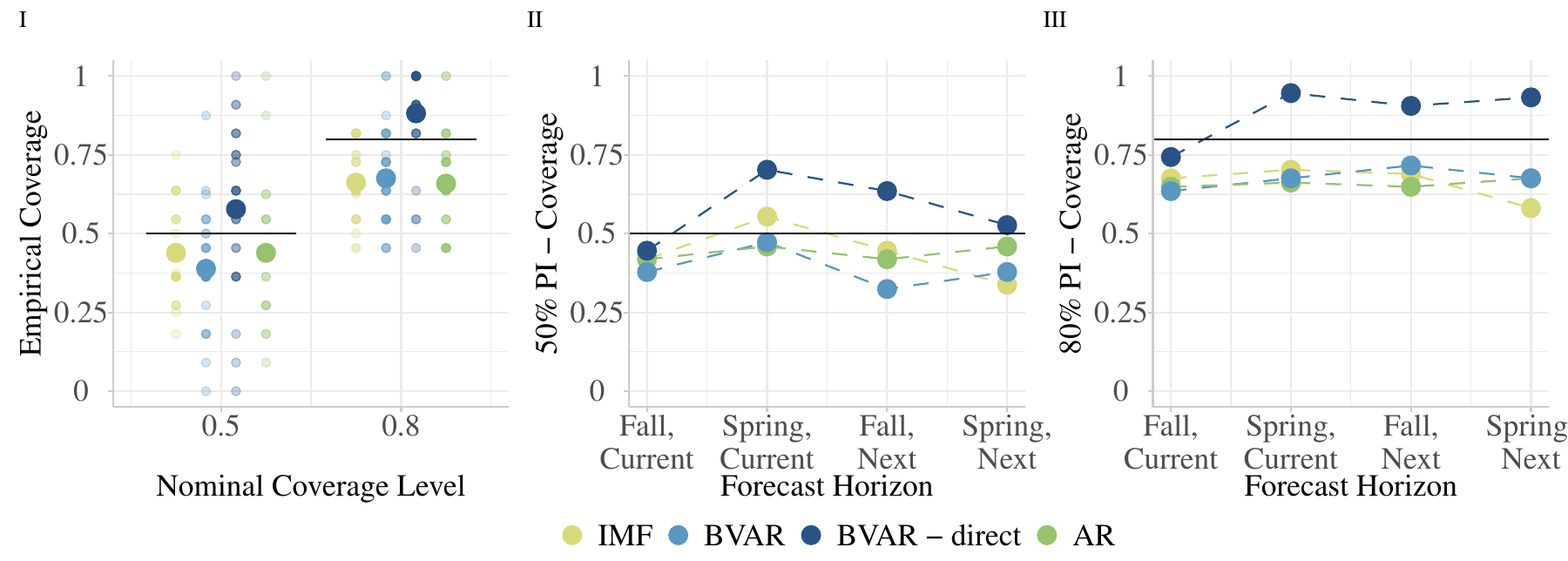}
    \caption{Inflation} 
    \label{fig:oecd_coverage_pcpi_pch_rollingwindow}
  \end{subfigure}
  
  \vfill  
  \begin{subfigure}{\linewidth}
    \centering
    \includegraphics[width=1.01\linewidth]{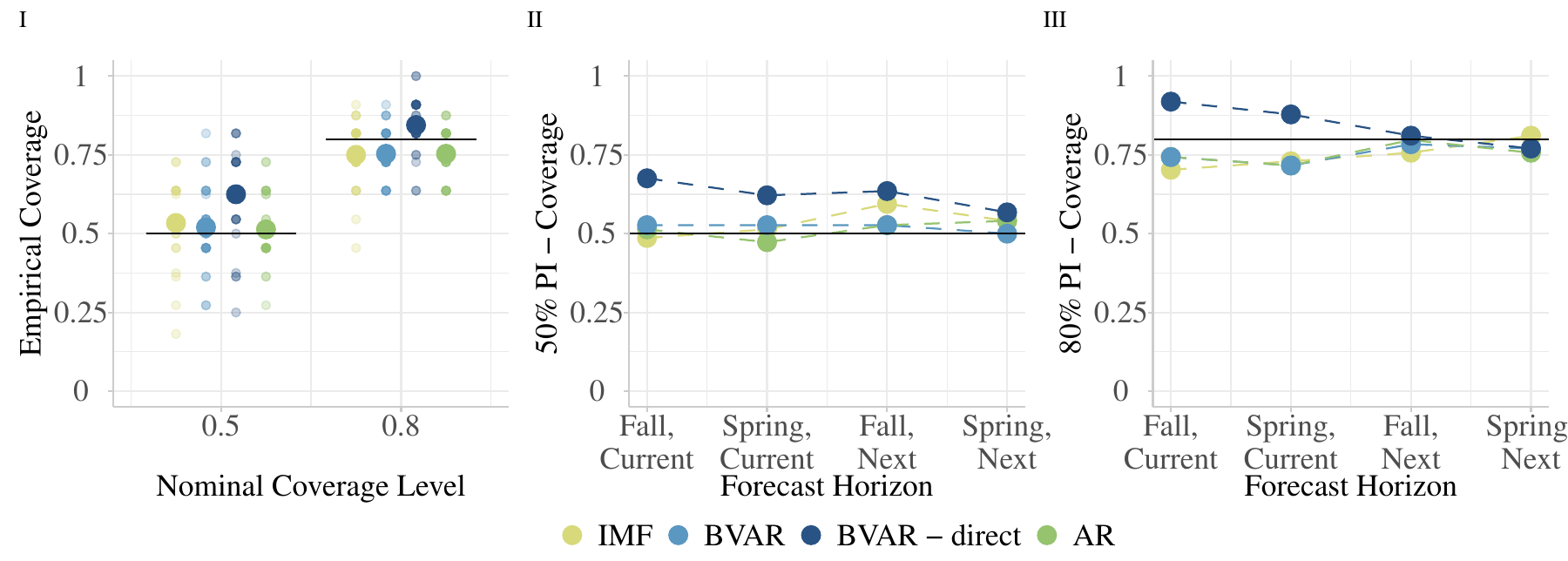}
    \caption{GDP growth} 
    \label{fig:oecd_coverage_ngdp_rpch_rollingwindow}
  \end{subfigure}
  \caption{Same as Figure \ref{fig:ho_coverage_absolute_rollingwindow}, but using OECD truth values as the outcome data.} \label{fig:ho_oecd_coverage_rollingwindow}
\end{figure}

\clearpage

\subsection{Details on Additional Benchmark Models}
\label{sec:details_bench}

\subsubsection{AR($p$)}

In addition to the parsimonious AR(1) specification from the main text, we consider choosing the AR lag length $p$ based on the \cite{Schwarz1978} information criterion as stated in Equation 4.3.9 of \cite{Luetkepohl2005} and discussed in Section 4.3 of the latter reference. While we allow for a maximal lag order of eight, the largest order chosen in practice is six. For inflation, lag orders between three and five are most common, whereas choices for GDP are clearly smaller, with one being the most popular and three being the maximal choice. 

\subsubsection{AR-annual and ARX-annual}

Here we describe two autoregressive specifications that are fitted to the annual predictand of interest. By comparison to the benchmark models based on quarterly (OECD) data, these annual specifications are more closely aligned with the prediction task at hand, and use a smaller (albeit less noisy) number of annual observations. 

The AR-annual specification is simply an AR(1) model for the annual outcome data. It is given by 
\begin{align*}
    y_t = \nu + \alpha~y_{t-1} + u_t,
\end{align*}
where $y_t$ is the annual observation (GDP growth rate or inflation rate) of interest, $\nu$ is an intercept term, $-1 < \alpha < 1$ is an autoregressive term and $u_t$ is the model's error term that is assumed to be independently normally distributed with variance $\sigma^2$. We estimate $\nu$ and $\alpha$ via ordinary least squares and $\sigma^2$ by the mean of squared residuals. 

To describe ARX-annual, let $\hat{y}_{\text{IMF},t,\text{FC}}$ denote the IMF point forecast for target year $t$ and target variable $y$, issued in fall of the same year (forecast horizon "Fall, Current"). Similarly, $\hat{y}_{\text{IMF},t,\text{SC}}$, $\hat{y}_{\text{IMF},t,\text{FN}}$ and $\hat{y}_{\text{IMF},t,\text{SN}}$ respectively denote the point forecast for target year $t$, made in spring of the same year (forecast horizon "Spring, Current"), fall of the previous year (forecast horizon "Fall, Next") and spring of the previous year (forecast horizon "Spring, Next"). Note thus again that $t$ refers to the target year: for instance, $\hat{y}_{\text{IMF},t,\text{FC}}$ is issued during the same year (year $t$), while $\hat{y}_{\text{IMF},t,\text{FN}}$ is issued in year $t-1$. Both issue forecasts for the same target year. 

The model equation for the ARX-annual model for \textit{same-year forecasts} with forecast origin fall then is
\begin{align*}
    y_t = \nu + \alpha y_{t-1} + \gamma~\hat{y}_{\text{IMF},t,\text{FC}} + u_t,
\end{align*}
where $u_t$ is an error term that we assume to be independently Gaussian. In spring, analogously,
\begin{align*}
    y_t = \nu + \alpha y_{t-1} + \gamma~\hat{y}_{\text{IMF},t,\text{SC}} + u_t.
\end{align*}

For \textit{next-year forecasts}, we similarly use the respective IMF point forecast for the target year as an additional predictor in the ARX-model. We additionally augment the model with the IMF point forecast for the current year, such that the model has knowledge of an estimate of the outcome in the current year.\footnote{as it does in the AR-annual multi-step predictions, where the model's estimate for the current year is "fed back" into the model when making predictions for the next year.}

The model equation for \textit{next-year forecasts} made in fall then is
\begin{align*}
    y_{t+1} = \nu + \alpha~y_{t-1} + \gamma_1~\hat{y}_{\text{IMF},{t+1},\text{FN}} + \gamma_2~\hat{y}_{\text{IMF},t,\text{FC}} + u_{t+1}.
\end{align*}
Note again that $\hat{y}_{\text{IMF},{t+1},\text{FN}}$ and $\hat{y}_{\text{IMF},t,\text{FC}}$ have the same forecast origin, that is, they are issued simultaneously in Fall during year $t$. In spring, the model equation is, analogously, 
\begin{align*}
    y_{t+1} = \nu + \alpha~y_{t-1} + \gamma_1~\hat{y}_{\text{IMF},{t+1},\text{SN}} + \gamma_2~\hat{y}_{\text{IMF},t,\text{SC}} + u_{t+1}.
\end{align*}

We estimate the four ARX-annual specifications separately across horizons, using ordinary least squares. Furthermore, we use the mean of squared residuals to estimate the error term variance. 

\subsubsection{BVAR-CISS and BVAR-Mix}

The BVAR-CISS model employs the constant parameter specification used in the main analysis, but adds the CISS index of financial stress \citep{HolloEtAl2012} as a third system variable, in addition to GDP growth and inflation. This specification is motivated by a recent strand of literature \citep[e.g.][]{AdrianEtAl2019} that discusses the predictive content of financial conditions on GDP growth and other macroeconomic variables. See Section \ref{sec:bmdata} of the supplement for details on the CISS data. 

As discussed in Section \ref{sec:bench}, the BVAR-Mix specification uses a time-varying parameter specification until the forecast year 2019, and switches to a constant parameter specification from 2020 onward.

\subsubsection{Simple Ensemble}

In order to briefly assess the potential of forecast combination in our context, we consider an equally-weighted quantile-wise average of the IMF, AR and BVAR models used in our main analysis. That is, at a given quantile level $q \in (0,1),$ the combined forecast is given by the simple mean of the IMF, AR and BVAR quantile predictions at level $q$. See e.g. \cite{LichtendahlEtAl2013} for further information and context on quantile based forecast combination.

\newpage
\subsection{Additional Forecast Intervals and Metrics}\label{sec:results_bench}
\begin{figure}[h!]
    \centering
    \includegraphics[width=0.95\linewidth]{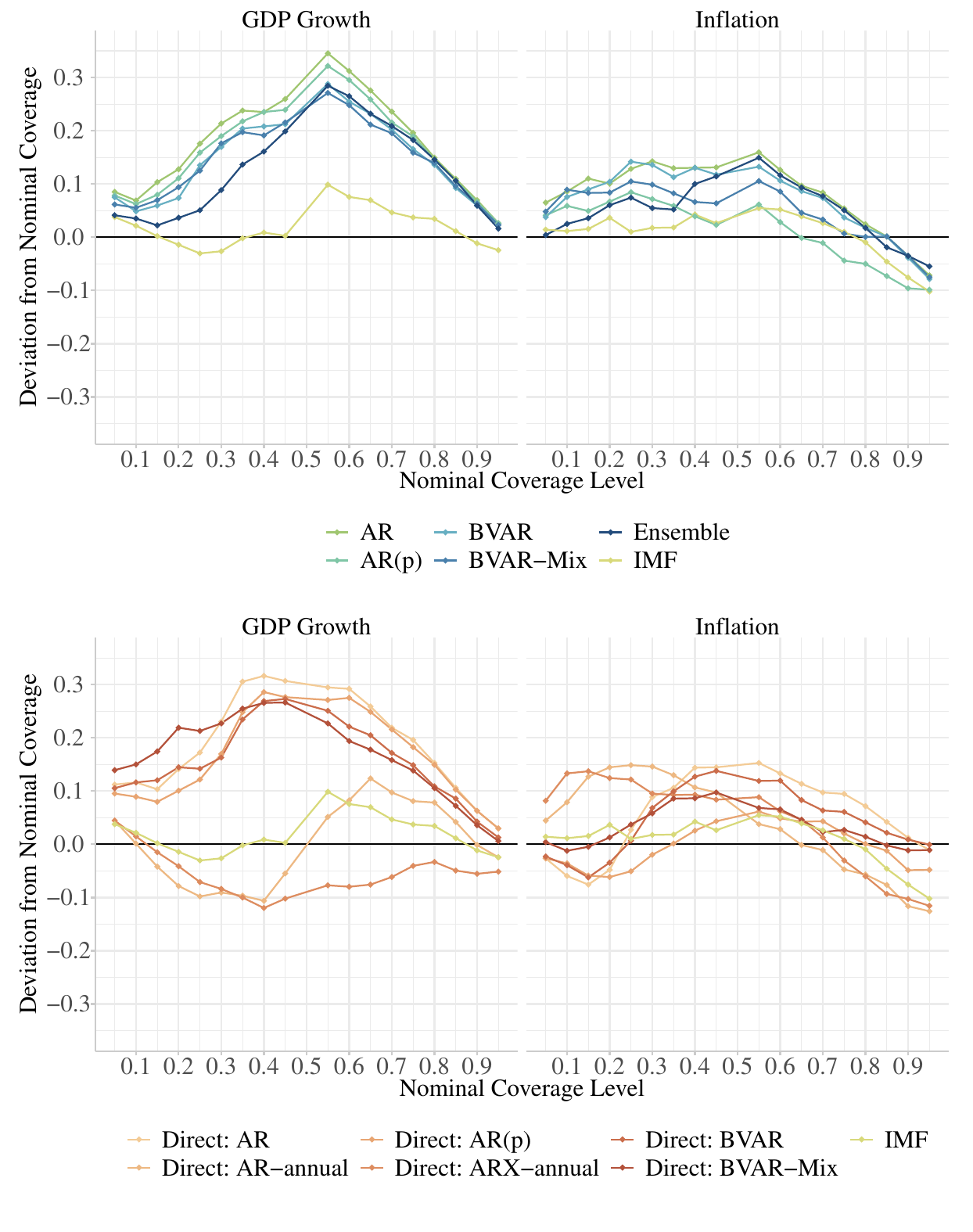}
    \caption{Empirical quantile coverage deviation in evaluation period 2013-2023, for quantile forecasts based on respective point forecast errors (top panel) and direct quantile forecasts (bottom panel), where forecast quantiles are taken from the respective model's parametric forecast distribution. Note that ``IMF" is generally based on point forecast errors, and is shown in both panels to enable comparison. Empirical quantile coverage is defined as the proportion of observations that fall below the forecast quantiles associated with a given nominal level. Coverage deviation is then computed as the empirical coverage level minus the nominal coverage level. Coverage levels are aggregated over the country and horizon dimension, and reported separately for the two targets (left: GDP growth, right: inflation). The horizontal line indicates nominal, that is, optimal coverage deviation levels.}
    \label{fig:extcis_qucvg}
\end{figure}
\newgeometry{left=2cm,right=2cm, bottom = 3.5cm, top = 3.5cm}
\begin{landscape}
\begingroup
		\renewcommand{\arraystretch}{1.1}

\begin{table}[!h]
\caption{Score values for current-year horizons, for extended set of benchmarks and extended set of forecast intervals}
\label{tab:extcis_currentyr}
\centering
\begin{tabular}{lp{4.5cm}p{1.3cm}p{1.3cm}p{1.3cm}p{1.3cm}p{1.95cm}p{1.3cm}p{1.3cm}p{1.3cm}p{1.3cm}p{1.3cm}}
\multicolumn{2}{c}{ } & \multicolumn{5}{c}{{Inflation\hspace*{10mm}}} & \multicolumn{5}{c}{{GDP Growth}} \\
 &  & $\text{CRPS}^{1)}$ & $\text{IS}_{W}^{2)}$ & $\text{IS}_{W,b}^{2)}$ & $\text{IS}_{U}^{2)}$ & $\text{IS}_{\text{CISS}}^{2)}$ & $\text{CRPS}^{1)}$ & $\text{IS}_{W}^{2)}$ & $\text{IS}_{W,b}^{2)}$ & $\text{IS}_{U}^{2)}$ & $\text{IS}_{\text{CISS}}^{2)}$\\
\midrule
\cellcolor{gray!35}{} & \cellcolor{gray!35}{IMF} & \cellcolor{gray!35}{\textbf{0.103}} & \cellcolor{gray!35}{\textbf{0.106}} & \cellcolor{gray!35}{\textbf{0.093}} & \cellcolor{gray!35}{\textbf{0.522}} & \cellcolor{gray!35}{\textbf{0.089}} & \cellcolor{gray!35}{0.292} & \cellcolor{gray!35}{0.298} & \cellcolor{gray!35}{0.269} & \cellcolor{gray!35}{1.561} & \cellcolor{gray!35}{0.255}\\
\parbox[t]{2mm}{\multirow{13}{*}{\rotatebox[origin=c]{90}{\hspace{5mm}Fall, Current}}}
 & AR$^{3)}$ & 0.195 & 0.198 & 0.177 & 1.034 & 0.214 & 0.337 & 0.344 & 0.298 & 1.697 & 0.302\\
 & AR($p$)$^{3)}$ & 0.183 & 0.185 & 0.166 & 0.978 & 0.203 & 0.333 & 0.34 & 0.296 & 1.688 & 0.3\\
 & BVAR$^{4)}$ & 0.2 & 0.203 & 0.183 & 1.053 & 0.219 & 0.326 & 0.334 & 0.285 & 1.664 & 0.288\\
 & BVAR-Mix$^{4)}$ & 0.198 & 0.201 & 0.182 & 1.042 & 0.218 & 0.323 & 0.33 & 0.284 & 1.642 & 0.287\\
 & Direct$^{5)}$: AR$^{3)}$ & 0.195 & 0.196 & 0.182 & 1.048 & 0.217 & 0.332 & 0.334 & 0.303 & 1.713 & 0.303\\
 & Direct$^{5)}$: AR($p$)$^{3)}$ & 0.181 & 0.183 & 0.169 & 0.981 & 0.203 & 0.326 & 0.328 & 0.298 & 1.684 & 0.3\\
 & Direct$^{5)}$: AR-annual$^{3)}$ & 0.984 & 0.988 & 0.899 & 5.179 & 0.987 & 1.604 & 1.613 & 1.56 & 8.995 & 1.678\\
 & Direct$^{5)}$: ARX-annual$^{3)}$ & 0.133 & 0.134 & 0.121 & 0.653 & 0.121 & 0.321 & 0.324 & 0.297 & 1.67 & 0.301\\
 & Direct$^{5)}$: BVAR$^{4)}$ & 0.2 & 0.201 & 0.184 & 1.069 & 0.221 & 0.328 & 0.33 & 0.303 & 1.735 & 0.305\\
 & Direct$^{5)}$: BVAR-CISS$^{4)}$ & -- & -- & -- & -- & 0.244 & -- & -- & -- & -- & 0.344\\
 & Direct$^{5)}$: BVAR-Mix$^{4)}$ & 0.2 & 0.201 & 0.185 & 1.07 & 0.224 & 0.335 & 0.336 & 0.311 & 1.798 & 0.311\\
\cellcolor{gray!15}{} & \cellcolor{gray!15}{Simple Ensemble$^{6)}$} & \cellcolor{gray!15}{0.145} & \cellcolor{gray!15}{0.147} & \cellcolor{gray!15}{0.132} & \cellcolor{gray!15}{0.752} & \cellcolor{gray!15}{0.156} & \cellcolor{gray!15}{\textbf{0.23}} & \cellcolor{gray!15}{\textbf{0.237}} & \cellcolor{gray!15}{\textbf{0.208}} & \cellcolor{gray!15}{\textbf{1.194}} & \cellcolor{gray!15}{\textbf{0.201}}\\
\addlinespace
\cellcolor{gray!35}{} & \cellcolor{gray!35}{IMF} & \cellcolor{gray!35}{0.419} & \cellcolor{gray!35}{0.425} & \cellcolor{gray!35}{0.387} & \cellcolor{gray!35}{2.25} & \cellcolor{gray!35}{0.425} & \cellcolor{gray!35}{\textbf{0.506}} & \cellcolor{gray!35}{\textbf{0.518}} & \cellcolor{gray!35}{\textbf{0.466}} & \cellcolor{gray!35}{2.64} & \cellcolor{gray!35}{0.496}\\
\parbox[t]{2mm}{\multirow{13}{*}{\rotatebox[origin=c]{90}{\hspace{5mm}Spring, Current}}}
 & AR$^{3)}$ & 0.454 & 0.46 & 0.405 & 2.319 & 0.441 & 0.816 & 0.829 & 0.742 & 4.292 & 0.699\\
 & AR($p$)$^{3)}$ & 0.464 & 0.47 & 0.438 & 2.401 & 0.473 & 0.739 & 0.751 & 0.676 & 3.91 & 0.649\\
 & BVAR$^{4)}$ & 0.455 & 0.461 & 0.41 & 2.334 & 0.45 & 0.716 & 0.728 & 0.651 & 3.763 & 0.606\\
 & BVAR-Mix$^{4)}$ & 0.458 & 0.465 & 0.412 & 2.365 & 0.453 & 0.697 & 0.709 & 0.637 & 3.66 & 0.594\\
 & Direct$^{5)}$: AR$^{3)}$ & 0.453 & 0.458 & 0.419 & 2.341 & 0.444 & 0.796 & 0.802 & 0.739 & 4.216 & 0.696\\
 & Direct$^{5)}$: AR($p$)$^{3)}$ & 0.462 & 0.466 & 0.43 & 2.386 & 0.469 & 0.72 & 0.726 & 0.677 & 3.83 & 0.654\\
 & Direct$^{5)}$: AR-annual$^{3)}$ & 0.983 & 0.988 & 0.899 & 5.178 & 0.987 & 1.607 & 1.617 & 1.563 & 9.008 & 1.681\\
 & Direct$^{5)}$: ARX-annual$^{3)}$ & 0.46 & 0.462 & 0.424 & 2.402 & 0.476 & 0.514 & \textbf{0.518} & 0.468 & \textbf{2.615} & 0.511\\
 & Direct$^{5)}$: BVAR$^{4)}$ & 0.456 & 0.462 & 0.426 & 2.358 & 0.458 & 0.69 & 0.695 & 0.637 & 3.625 & 0.59\\
 & Direct$^{5)}$: BVAR-CISS$^{4)}$ & -- & -- & -- & -- & 0.502 & -- & -- & -- & -- & 0.671\\
 & Direct$^{5)}$: BVAR-Mix$^{4)}$ & 0.429 & 0.433 & 0.384 & 2.169 & 0.411 & 0.667 & 0.671 & 0.611 & 3.502 & 0.569\\
\cellcolor{gray!15}{} & \cellcolor{gray!15}{Simple Ensemble$^{6)}$} & \cellcolor{gray!15}{\textbf{0.389}} & \cellcolor{gray!15}{\textbf{0.395}} & \cellcolor{gray!15}{\textbf{0.351}} & \cellcolor{gray!15}{\textbf{1.998}} & \cellcolor{gray!15}{\textbf{0.384}} & \cellcolor{gray!15}{0.559} & \cellcolor{gray!15}{0.571} & \cellcolor{gray!15}{0.508} & \cellcolor{gray!15}{2.893} & \cellcolor{gray!15}{\textbf{0.489}}\\
\bottomrule

	\multicolumn{12}{l}{\scriptsize
	${1)}$ Sample-based Continuous Ranked Probability Score, see \cite{jordan19}.}\\
	\multicolumn{12}{l}{\parbox{\linewidth}{\scriptsize
	${2)}$ Different Versions of the Interval Score. $\text{IS}_{W}$ represents a weighted sum of interval scores at extended set of levels $\tau \in \{0.1, 0.2, ... 0.9\}$, with weights as given by \cite{bracher2020}. $\text{IS}_{W,b}$ represents the analogous weighted sum of interval scores with forecast intervals $\tau \in \{0.5, 0.8\}$ from main analysis. $\text{IS}_{U}$ represents a simple equally-weighted average of interval scores at extended set of levels $\tau \in \{0.1, 0.2, ... 0.9\}$. $\text{IS}_{\text{CISS}}$ represents the weighted sum of interval scores with forecast intervals $\tau \in \{0.5, 0.8\}$, for a subset of the data that excludes Canada and Japan due to availability of the CISS-data.}}\\[-0.3em]
    \multicolumn{12}{l}{\scriptsize
	${3)}$ Different autoregressive model specifications. `AR' refers to the AR model from main analysis. For more details, please refer to Section \ref{sec:details_bench} in the supplement.}\\[-0.5em]
	\multicolumn{12}{l}{\scriptsize
	${4)}$ Different Bayesian vector autoregressive model specifications. `BVAR' refers to the BVAR model from main analysis. For more details, please refer to Section \ref{sec:details_bench} in the supplement.}\\[-0.5em]
	\multicolumn{12}{l}{\scriptsize
	${5)}$ Forecast quantiles are taken directly from the parametric forecast distribution of the respective model.}\\[-0.5em]
	\multicolumn{12}{l}{\scriptsize
	${6)}$ Quantile-wise simple average of models from main analysis, that is, the 'IMF', 'AR', and 'BVAR' forecasts. }
\end{tabular}
\end{table}
\endgroup
\end{landscape}
\newpage
\begin{landscape}
\begingroup
		\renewcommand{\arraystretch}{1.1}
\begin{table}[!h]
\caption{Score values for next-year horizons, for extended set of benchmarks and extended set of forecast intervals}
\label{tab:extcis_nextyr}
\centering
\begin{tabular}{lp{4.5cm}p{1.3cm}p{1.3cm}p{1.3cm}p{1.3cm}p{1.95cm}p{1.3cm}p{1.3cm}p{1.3cm}p{1.3cm}p{1.3cm}}
\multicolumn{2}{c}{ } & \multicolumn{5}{c}{{Inflation\hspace*{15mm}}} & \multicolumn{5}{c}{{GDP Growth}} \\
 &  & $\text{CRPS}^{1)}$ & $\text{IS}_{W}^{2)}$ & $\text{IS}_{W,b}^{2)}$ & $\text{IS}_{U}^{2)}$ & $\text{IS}_{\text{CISS}}^{2)}$ & $\text{CRPS}^{1)}$ & $\text{IS}_{W}^{2)}$ & $\text{IS}_{W,b}^{2)}$ & $\text{IS}_{U}^{2)}$ & $\text{IS}_{\text{CISS}}^{2)}$\\
\midrule
\cellcolor{gray!35}{} & \cellcolor{gray!35}{IMF} & \cellcolor{gray!35}{0.914} & \cellcolor{gray!35}{0.925} & \cellcolor{gray!35}{0.825} & \cellcolor{gray!35}{4.863} & \cellcolor{gray!35}{0.899} & \cellcolor{gray!35}{\textbf{1.173}} & \cellcolor{gray!35}{\textbf{1.189}} & \cellcolor{gray!35}{\textbf{1.128}} & \cellcolor{gray!35}{\textbf{6.475}} & \cellcolor{gray!35}{\textbf{1.188}}\\
\parbox[t]{2mm}{\multirow{13}{*}{\rotatebox[origin=c]{90}{\hspace{5mm}Fall, Next}}}
 & AR$^{3)}$ & 1.056 & 1.071 & 0.904 & 5.08 & 0.977 & 1.494 & 1.514 & 1.367 & 7.944 & 1.368\\
 & AR($p$)$^{3)}$ & 0.955 & 0.968 & 0.851 & 4.903 & 0.9 & 1.427 & 1.446 & 1.32 & 7.689 & 1.328\\
 & BVAR$^{4)}$ & 1.006 & 1.02 & 0.857 & 4.883 & 0.954 & 1.368 & 1.387 & 1.289 & 7.437 & 1.313\\
 & BVAR-Mix$^{4)}$ & 0.876 & 0.888 & \textbf{0.758} & 4.332 & 0.854 & 1.334 & 1.352 & 1.26 & 7.269 & 1.287\\
 & Direct$^{5)}$: AR$^{3)}$ & 1.037 & 1.049 & 0.91 & 5.052 & 0.973 & 1.487 & 1.497 & 1.381 & 7.978 & 1.383\\
 & Direct$^{5)}$: AR($p$)$^{3)}$ & 0.93 & 0.938 & 0.844 & 4.689 & 0.88 & 1.417 & 1.427 & 1.34 & 7.701 & 1.349\\
 & Direct$^{5)}$: AR-annual$^{3)}$ & 1.296 & 1.301 & 1.194 & 6.913 & 1.354 & 1.467 & 1.476 & 1.424 & 8.127 & 1.524\\
 & Direct$^{5)}$: ARX-annual$^{3)}$ & 1.056 & 1.06 & 0.976 & 5.613 & 1.093 & 1.482 & 1.49 & 1.381 & 7.969 & 1.523\\
 & Direct$^{5)}$: BVAR$^{4)}$ & 0.995 & 1.007 & 0.887 & 4.903 & 0.952 & 1.361 & 1.37 & 1.29 & 7.458 & 1.31\\
 & Direct$^{5)}$: BVAR-CISS$^{4)}$ & -- & -- & -- & -- & 1.05 & -- & -- & -- & -- & 1.871\\
 & Direct$^{5)}$: BVAR-Mix$^{4)}$ & \textbf{0.857} & \textbf{0.866} & 0.771 & \textbf{4.215} & \textbf{0.851} & 1.338 & 1.346 & 1.27 & 7.428 & 1.286\\
\cellcolor{gray!15}{} & \cellcolor{gray!15}{Simple Ensemble$^{6)}$} & \cellcolor{gray!15}{0.93} & \cellcolor{gray!15}{0.943} & \cellcolor{gray!15}{0.817} & \cellcolor{gray!15}{4.644} & \cellcolor{gray!15}{0.896} & \cellcolor{gray!15}{1.286} & \cellcolor{gray!15}{1.304} & \cellcolor{gray!15}{1.213} & \cellcolor{gray!15}{7.011} & \cellcolor{gray!15}{1.242}\\
\addlinespace
\cellcolor{gray!35}{} & \cellcolor{gray!35}{IMF} & \cellcolor{gray!35}{1.108} & \cellcolor{gray!35}{1.119} & \cellcolor{gray!35}{1.013} & \cellcolor{gray!35}{5.895} & \cellcolor{gray!35}{1.123} & \cellcolor{gray!35}{\textbf{1.25}} & \cellcolor{gray!35}{\textbf{1.271}} & \cellcolor{gray!35}{\textbf{1.204}} & \cellcolor{gray!35}{\textbf{6.878}} & \cellcolor{gray!35}{\textbf{1.267}}\\
\parbox[t]{2mm}{\multirow{13}{*}{\rotatebox[origin=c]{90}{\hspace{5mm}Spring, Next}}}
 & AR$^{3)}$ & 1.377 & 1.396 & 1.142 & 6.349 & 1.195 & 1.769 & 1.791 & 1.587 & 9.22 & 1.6\\
 & AR($p$)$^{3)}$ & 1.113 & 1.126 & 0.975 & 5.683 & 1.057 & 1.757 & 1.781 & 1.608 & 9.347 & 1.667\\
 & BVAR$^{4)}$ & 1.346 & 1.363 & 1.122 & 6.304 & 1.215 & 1.686 & 1.708 & 1.571 & 9.124 & 1.635\\
 & BVAR-Mix$^{4)}$ & 1.126 & 1.141 & 0.962 & 5.408 & 1.06 & 1.635 & 1.656 & 1.535 & 8.874 & 1.607\\
 & Direct$^{5)}$: AR$^{3)}$ & 1.35 & 1.365 & 1.131 & 6.367 & 1.185 & 1.763 & 1.775 & 1.601 & 9.183 & 1.609\\
 & Direct$^{5)}$: AR($p$)$^{3)}$ & \textbf{1.095} & \textbf{1.106} & 1.002 & 5.63 & 1.061 & 1.719 & 1.731 & 1.579 & 9.111 & 1.646\\
 & Direct$^{5)}$: AR-annual$^{3)}$ & 1.296 & 1.301 & 1.194 & 6.912 & 1.354 & 1.467 & 1.477 & 1.424 & 8.128 & 1.525\\
 & Direct$^{5)}$: ARX-annual$^{3)}$ & 1.241 & 1.245 & 1.155 & 6.68 & 1.319 & 1.396 & 1.404 & 1.33 & 7.611 & 1.418\\
 & Direct$^{5)}$: BVAR$^{4)}$ & 1.336 & 1.352 & 1.133 & 6.39 & 1.194 & 1.651 & 1.661 & 1.533 & 8.848 & 1.601\\
 & Direct$^{5)}$: BVAR-CISS$^{4)}$ & -- & -- & -- & -- & 1.103 & -- & -- & -- & -- & 1.876\\
 & Direct$^{5)}$: BVAR-Mix$^{4)}$ & \textbf{1.095} & 1.107 & \textbf{0.942} & \textbf{5.254} & \textbf{1.02} & 1.611 & 1.62 & 1.501 & 8.729 & 1.575\\
\cellcolor{gray!15}{} & \cellcolor{gray!15}{Simple Ensemble$^{6)}$} & \cellcolor{gray!15}{1.18} & \cellcolor{gray!15}{1.196} & \cellcolor{gray!15}{0.993} & \cellcolor{gray!15}{5.617} & \cellcolor{gray!15}{1.069} & \cellcolor{gray!15}{1.509} & \cellcolor{gray!15}{1.53} & \cellcolor{gray!15}{1.403} & \cellcolor{gray!15}{8.122} & \cellcolor{gray!15}{1.455}\\
\bottomrule

	\multicolumn{12}{l}{\scriptsize
	${1)}$ Sample-based Continuous Ranked Probability Score, see \cite{jordan19}.}\\
	\multicolumn{12}{l}{\parbox{\linewidth}{\scriptsize
	${2)}$ Different Versions of the Interval Score. $\text{IS}_{W}$ represents a weighted sum of interval scores at extended set of levels $\tau \in \{0.1, 0.2, ... 0.9\}$, with weights as given by \cite{bracher2020}. $\text{IS}_{W,b}$ represents the analogous weighted sum of interval scores with forecast intervals $\tau \in \{0.5, 0.8\}$ from main analysis. $\text{IS}_{U}$ represents a simple equally-weighted average of interval scores at extended set of levels $\tau \in \{0.1, 0.2, ... 0.9\}$. $\text{IS}_{\text{CISS}}$ represents the weighted sum of interval scores with forecast intervals $\tau \in \{0.5, 0.8\}$, for a subset of the data that excludes Canada and Japan due to availability of the CISS-data.}}\\[-0.3em]
    \multicolumn{12}{l}{\scriptsize
	${3)}$ Different autoregressive model specifications. `AR' refers to the AR model from main analysis. For more details, please refer to Section \ref{sec:details_bench} in the supplement.}\\[-0.5em]
	\multicolumn{12}{l}{\scriptsize
	${4)}$ Different Bayesian vector autoregressive model specifications. `BVAR' refers to the BVAR model from main analysis. For more details, please refer to Section \ref{sec:details_bench} in the supplement.}\\[-0.5em]
	\multicolumn{12}{l}{\scriptsize
	${5)}$ Forecast quantiles are taken directly from the parametric forecast distribution of the respective model.}\\[-0.5em]
	\multicolumn{12}{l}{\scriptsize
	${6)}$ Quantile-wise simple average of models from main analysis, that is, the 'IMF', 'AR', and 'BVAR' forecasts. }
\end{tabular}
\end{table}
\endgroup
\end{landscape}
\restoregeometry

\begin{table}[!h]
\centering
\caption{Individual interval scores, for extended set of forecast intervals}
\label{tab:extcis_isscores}
\begin{tabular}{p{0.8cm}lllllllll}
\toprule
\multicolumn{2}{c}{ } & \multicolumn{4}{c}{{Inflation}} & \multicolumn{4}{c}{{GDP Growth}} \\
\cmidrule(l{3pt}r{3pt}){3-6} \cmidrule(l{3pt}r{3pt}){7-10}
 &  & $\text{IS}_{30}^{1)}$ & $\text{IS}_{50}^{1)}$ & $\text{IS}_{80}^{1)}$ & $\text{IS}_{90}^{1)}$ & $\text{IS}_{30}^{1)}$ & $\text{IS}_{50}^{1)}$ & $\text{IS}_{80}^{1)}$ & $\text{IS}_{90}^{1)}$\\
\midrule
\parbox[t]{2mm}{\multirow{4}{*}{\rotatebox[origin=c]{90}{\parbox{2cm}{\centering Fall,\\Current}}}} & IMF & \textbf{0.38} & \textbf{0.46} & \textbf{0.74} & \textbf{0.92} & \textbf{0.99} & \textbf{1.28} & \textbf{2.3} & 3.34\\
 & AR & 0.67 & 0.84 & 1.48 & 2.32 & 1.21 & 1.48 & 2.36 & 3.19\\
 & BVAR & 0.69 & 0.87 & 1.49 & 2.29 & 1.17 & 1.45 & 2.36 & \textbf{3.18}\\
 & Direct$^{2)}$: BVAR & 0.68 & 0.86 & 1.54 & 2.43 & 1.13 & 1.43 & 2.49 & 3.78\\
\addlinespace

\parbox[t]{2mm}{\multirow{4}{*}{\rotatebox[origin=c]{90}{\parbox{2cm}{\centering Spring,\\Current}}}} & IMF & \textbf{1.43} & \textbf{1.81} & 3.26 & 5.11 & \textbf{1.74} & \textbf{2.25} & \textbf{3.96} & \textbf{5.12}\\
 & AR & 1.58 & 1.96 & \textbf{3.24} & 4.7 & 2.8 & 3.49 & 6.19 & 9.21\\
 & BVAR & 1.57 & 1.97 & 3.33 & 4.7 & 2.44 & 3.13 & 5.37 & 7.94\\
 & Direct$^{2)}$: BVAR & 1.64 & 2.06 & 3.35 & \textbf{4.47} & 2.4 & 3.03 & 5.16 & 7.68\\
\addlinespace

\parbox[t]{2mm}{\multirow{4}{*}{\rotatebox[origin=c]{90}{\parbox{2cm}{\centering Fall,\\Next}}}} & IMF & \textbf{3.12} & \textbf{3.87} & 6.94 & 11.1 & \textbf{3.79} & \textbf{5.11} & \textbf{9.73} & \textbf{15.39}\\
 & AR & 3.92 & 4.58 & 6.65 & 9.02 & 5.03 & 6.37 & 11.43 & 17.9\\
 & BVAR & 3.69 & 4.37 & \textbf{6.4} & 8.86 & 4.54 & 5.84 & 11.01 & 17.43\\
 & Direct$^{2)}$: BVAR & 3.62 & 4.39 & 6.76 & \textbf{8.27} & 4.52 & 5.9 & 11.04 & 17.46\\
\addlinespace
\parbox[t]{2mm}{\multirow{4}{*}{\rotatebox[origin=c]{90}{\parbox{2cm}{\centering Spring,\\Next}}}} & IMF & \textbf{3.75} & \textbf{4.74} & 8.4 & 13.64 & \textbf{4.11} & \textbf{5.48} & \textbf{10.37} & \textbf{16.05}\\
 & AR & 5.23 & 5.95 & \textbf{7.92} & \textbf{9.75} & 6.06 & 7.52 & 13.04 & 19.88\\
 & BVAR & 5.02 & 5.76 & 8 & 10.32 & 5.58 & 7.24 & 13.28 & 21.31\\
 & Direct$^{2)}$: BVAR & 5.03 & 5.7 & 8.41 & 10.63 & 5.59 & 7.14 & 12.82 & 19.8\\
\bottomrule \\[-1em] 
\multicolumn{10}{l}{\parbox{0.89\linewidth}{\scriptsize
	${1)}$ Different Versions of the Interval Score. $\text{IS}_{30}$, $\text{IS}_{50}$, $\text{IS}_{80}$  and $\text{IS}_{90}$ represent the interval score for the  $30\%$,  $50\%$, $80\%$ and $90\%$ intervals, respectively. Note that for the non-direct forecast intervals, IS values shown here are sometimes slightly different than those shown in Table \ref{tab:baseis5080} showing results from the main analysis, due to the PAVA-type correction algorithm correcting more instances than in the main analysis.}}\\
	\multicolumn{10}{l}{\scriptsize
	${2)}$ Forecast quantiles are taken directly from the parametric forecast distribution of the respective model.}
\end{tabular}
\end{table}

\newpage
\subsection{Results for Additional Countries}
\subsubsection{Inflation}
\begin{table}[!h]
\caption{Scores for evaluation on extended country set, for period 2013-2023 and forecast target inflation.}
\label{tab:extctry_pcpi_pch}
\centering
\begin{tabular}{lllllll}
\toprule
 &  & $\text{CRPS}^{1)}$ & $\text{IS}_{W,b}^{2)}$ & $\text{IS}_{U}^{2)}$ & $\text{IS}_{50}^{2)}$ & $\text{IS}_{80}^{2)}$\\
\midrule
\multirow{3}{*}{Fall, Current} & IMF & \textbf{0.224} & \textbf{0.183} & \textbf{1.136} & \textbf{0.927} & \textbf{1.344}\\

 & Direct$^{3)}$: AR-annual & 1.469 & 1.202 & 7.639 & 5.848 & 9.429\\
 & Direct$^{3)}$: ARX-annual & 0.344 & 0.266 & 1.663 & 1.326 & 2\\
\addlinespace
\multirow{3}{*}{Spring, Current} & IMF & \textbf{0.674} & \textbf{0.59} & \textbf{3.807} & \textbf{2.789} & \textbf{4.825}\\

 & Direct$^{3)}$: AR-annual & 1.47 & 1.204 & 7.643 & 5.858 & 9.428\\
 & Direct$^{3)}$: ARX-annual & 0.892 & 0.745 & 4.746 & 3.606 & 5.886\\
\addlinespace
\multirow{3}{*}{Fall, Next} & IMF & \textbf{1.251} & \textbf{1.068} & \textbf{6.844} & \textbf{5.119} & \textbf{8.569}\\

 & Direct$^{3)}$: AR-annual & 1.718 & 1.403 & 8.879 & 6.868 & 10.89\\
 & Direct$^{3)}$: ARX-annual & 1.68 & 1.42 & 9.038 & 6.878 & 11.197\\
\addlinespace
\multirow{3}{*}{Spring, Next} & IMF & \textbf{1.51} & \textbf{1.289} & \textbf{8.28} & \textbf{6.152} & \textbf{10.409}\\

 & Direct$^{3)}$: AR-annual & 1.716 & 1.402 & 8.874 & 6.858 & 10.889\\
 & Direct$^{3)}$: ARX-annual & 1.814 & 1.544 & 9.905 & 7.387 & 12.423\\
\bottomrule
\multicolumn{7}{l}{\scriptsize
	${1)}$ Sample-based Continuous Ranked Probability Score, see \cite{jordan19}.}\\
	\multicolumn{7}{l}{\parbox{\linewidth}{\scriptsize
	${2)}$ Different Versions of the Interval Score. $\text{IS}_{50}$ and $\text{IS}_{80}$ represent the Interval Score for the $50\%$ and 	$80\%$ intervals, respectively. $\text{IS}_{W,b}$ represents a weighted sum of interval scores from quantile forecasts at base 	levels from main analysis $\tau \in \{0.5, 0.8\}$, with weights as given by \cite{bracher2020}. $\text{IS}_{U}$ represents a simple equally-weighted average of interval scores, here at base levels $\tau \in \{0.5, 0.8\}$. }}\\
	\multicolumn{7}{l}{\scriptsize
	${3)}$ Forecast quantiles are taken directly from the parametric forecast distribution of the respective model.}
\end{tabular}
\end{table}

\vspace{0.5cm}
\begin{figure}[h!]
    \centering
    \includegraphics[width=1.01\linewidth]{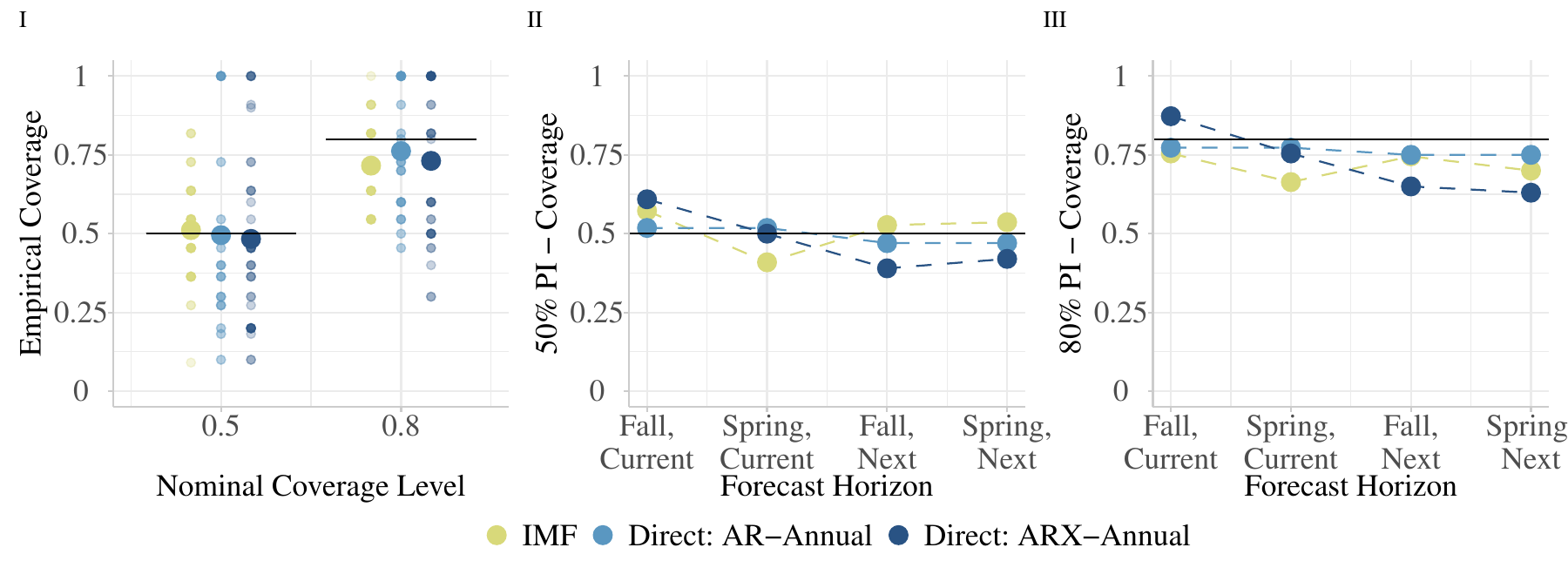}
    \caption{Same as Figure \ref{fig:ho_coverage_absolute_rollingwindow}, but on extended country set, and for Inflation only.}
    \label{fig:cvg_extctry_pcpi_pch}
\end{figure}
\newpage
\subsubsection{GDP Growth}
\begin{table}[!h]
\caption{Scores for evaluation on extended country set, for period 2013-2023 and forecast target GDP growth.}
\label{tab:extctry_ngdp_rpch}
\centering
\begin{tabular}{lllllll}
\toprule
 &  & $\text{CRPS}^{1)}$ & $\text{IS}_{W,b}^{2)}$ & $\text{IS}_{U}^{2)}$ & $\text{IS}_{50}^{2)}$ & $\text{IS}_{80}^{2)}$\\
\midrule
\multirow{3}{*}{Fall, Current} & IMF & \textbf{0.435} & \textbf{0.378} & \textbf{2.465} & \textbf{1.759} & \textbf{3.17}\\

 & Direct$^{3)}$: AR-annual & 1.923 & 1.655 & 10.674 & 7.835 & 13.513\\
 & Direct$^{3)}$: ARX-annual & 0.561 & 0.461 & 2.935 & 2.237 & 3.633\\
\addlinespace
\multirow{3}{*}{Spring, Current} & IMF & \textbf{0.849} & \textbf{0.732} & \textbf{4.697} & \textbf{3.499} & \textbf{5.895}\\

 & Direct$^{3)}$: AR-annual & 1.938 & 1.67 & 10.765 & 7.907 & 13.623\\
 & Direct$^{3)}$: ARX-annual & 1.135 & 0.96 & 6.162 & 4.582 & 7.741\\
\addlinespace
\multirow{3}{*}{Fall, Next} & IMF & \textbf{1.39} & \textbf{1.208} & \textbf{7.819} & \textbf{5.681} & \textbf{9.956}\\

 & Direct$^{3)}$: AR-annual & 1.892 & 1.607 & 10.28 & 7.718 & 12.842\\
 & Direct$^{3)}$: ARX-annual & 1.942 & 1.698 & 10.892 & 8.114 & 13.669\\
\addlinespace
\multirow{3}{*}{Spring, Next} & IMF & \textbf{1.572} & \textbf{1.344} & \textbf{8.651} & \textbf{6.38} & \textbf{10.921}\\

 & Direct$^{3)}$: AR-annual & 1.902 & 1.617 & 10.344 & 7.765 & 12.923\\
 & Direct$^{3)}$: ARX-annual & 1.986 & 1.73 & 11.162 & 8.184 & 14.139\\
\bottomrule
\multicolumn{7}{l}{\scriptsize
	${1)}$ Sample-based Continuous Ranked Probability Score, see \cite{jordan19}.}\\
	\multicolumn{7}{l}{\parbox{\linewidth}{\scriptsize
	${2)}$ Different Versions of the Interval Score. $\text{IS}_{50}$ and $\text{IS}_{80}$ represent the Interval Score for the $50\%$ and $80\%$ intervals, respectively. $\text{IS}_{W,b}$ represents a weighted sum of interval scores from quantile forecasts at base 	levels from main analysis $\tau \in \{0.5, 0.8\}$, with weights as given by \cite{bracher2020}. $\text{IS}_{U}$ represents a simple equally-weighted average of interval scores, here at base levels $\tau \in \{0.5, 0.8\}$. }}\\
	\multicolumn{7}{l}{\scriptsize
	${3)}$ Forecast quantiles are taken directly from the parametric forecast distribution of the respective model.}
\end{tabular}
\end{table}

\vspace{0.5cm}
\begin{figure}[h!]
    \centering
    \includegraphics[width=1.01\linewidth]{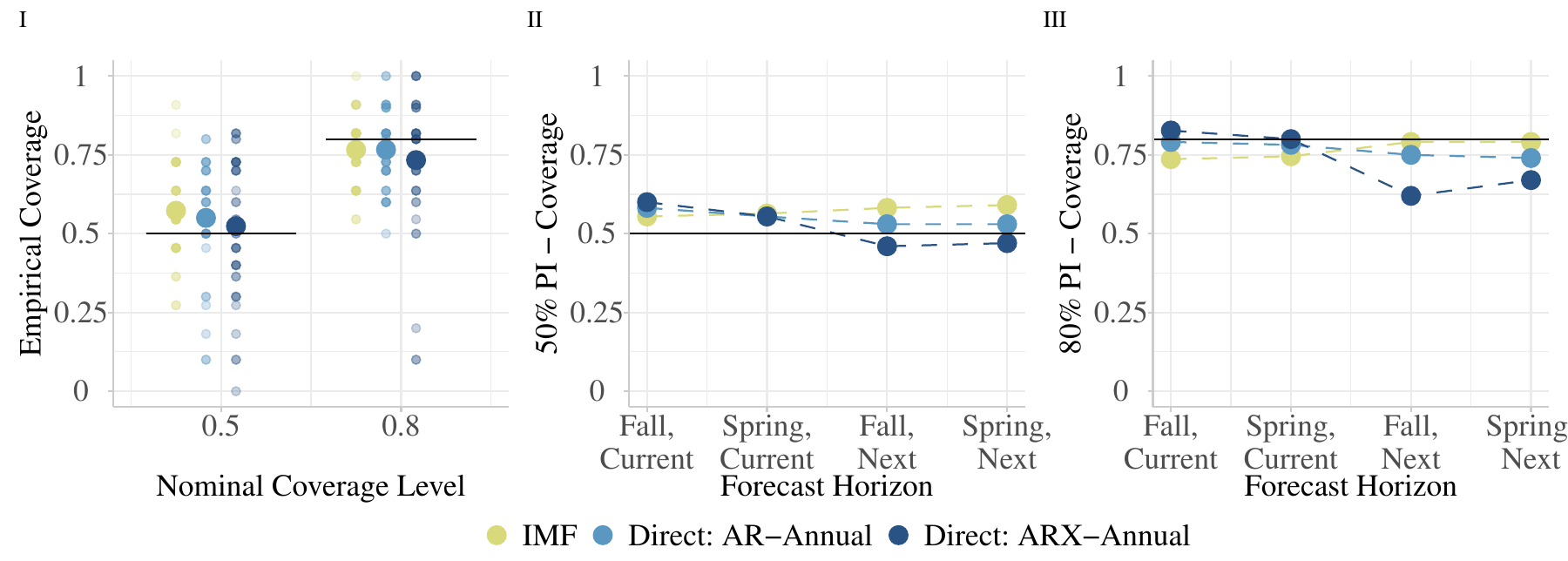}
    \caption{Same as Figure \ref{fig:ho_coverage_absolute_rollingwindow}, but on extended country set, and for GDP growth only.}
    \label{fig:cvg_extctry_ngdp_rpch}
\end{figure}
\newpage

\subsection{Forecasts During the Covid-19 Period}
\subsubsection{GDP Growth}
\begin{table}[!h]
\centering
\caption{Scores for Target Years 2020 - 2023 (COVID period), GDP Growth}
\label{tab:sens_cov19gdp}
\centering
\begin{tabular}[t]{p{0.8cm}lllllll}
\toprule
 &  & $\text{IS}_{W,b}^{1)}$ & $\text{IS}_{U}^{1)}$ & $\text{IS}_{50}^{1)}$ & $\text{IS}_{80}^{1)}$ & $\text{Dev}_{50}^{2)}$ & $\text{Dev}_{80}^{2)}$\\
\midrule
\parbox[t]{2mm}{\multirow{4}{*}{\rotatebox[origin=c]{90}{\parbox{2cm}{\centering Fall,\\Current}}}} & IMF & 0.495 & 3.278 & 2.229 & 4.328 & -26 & -36\\
 & AR & 0.418 & 2.663 & 2.019 & 3.306 & \textbf{-6} & \textbf{-4}\\
 & BVAR & \textbf{0.39} & \textbf{2.497} & \textbf{1.873} & \textbf{3.121} & \textbf{-6} & \textbf{-4}\\
 & Direct: BVAR & 0.432 & 2.821 & 2.002 & 3.64 & -30 & -48\\
\addlinespace
\parbox[t]{2mm}{\multirow{4}{*}{\rotatebox[origin=c]{90}{\parbox{2cm}{\centering Spring,\\Current}}}} & IMF & \textbf{0.765} & \textbf{4.997} & \textbf{3.543} & \textbf{6.45} & -26 & -24\\
 & AR & 1.379 & 9.11 & 6.238 & 11.983 & \textbf{-22} & -16\\
 & BVAR & 1.205 & 7.87 & 5.576 & 10.163 & -30 & -16\\
 & Direct: BVAR & 1.162 & 7.592 & 5.368 & 9.816 & \textbf{-22} & \textbf{-12}\\
\addlinespace
\parbox[t]{2mm}{\multirow{4}{*}{\rotatebox[origin=c]{90}{\parbox{2cm}{\centering Fall,\\Next}}}} & IMF & \textbf{2.508} & \textbf{16.652} & \textbf{11.234} & \textbf{22.071} & -34 & -44\\
 & AR & 2.698 & 17.89 & 12.118 & 23.663 & \textbf{-14} & \textbf{-16}\\
 & BVAR & 2.671 & 17.776 & 11.91 & 23.642 & -18 & -20\\
 & Direct: BVAR & 2.672 & 17.815 & 11.876 & 23.754 & \textbf{-14} & -20\\
\addlinespace
\parbox[t]{2mm}{\multirow{4}{*}{\rotatebox[origin=c]{90}{\parbox{2cm}{\centering Spring,\\Next}}}} & IMF & \textbf{2.511} & \textbf{16.549} & \textbf{11.414} & \textbf{21.684} & -38 & -40\\
 & AR & 3.119 & 20.586 & 14.136 & 27.037 & \textbf{-18} & \textbf{-36}\\
 & BVAR & 3.336 & 22.196 & 14.887 & 29.506 & -22 & \textbf{-36}\\
 & Direct: BVAR & 3.238 & 21.324 & 14.737 & 27.912 & -26 & \textbf{-36}\\
\bottomrule
\multicolumn{8}{l}{\parbox{0.81\linewidth}{\scriptsize
	${1)}$ Different Versions of the Interval Score. $\text{IS}_{50}$ and $\text{IS}_{80}$ represent the interval score for the $50\%$ and $80\%$ intervals, respectively. $\text{IS}_{W,b}$ represents the weighted sum of interval scores from quantile forecasts at base levels from main analysis $\tau \in \{0.5, 0.8\}$. $\text{IS}_{U}$ represents a simple equally-weighted average of interval scores at the same levels.}}\\
	\multicolumn{8}{l}{\parbox{0.81\linewidth}{\scriptsize
	${2)}$ Coverage deviation at nominal interval coverage levels $50\%$ and $80\%$. Negative values represent cases where the respective intervals covered fewer observations than the nominal level would indicate, and vice versa for positive values.}}
\end{tabular}
\end{table}

\newpage
\subsubsection{Inflation}
\begin{table}[!h]
\centering
\caption{Scores for Target Years 2020 - 2023 (COVID period), Inflation}
\label{tab:sens_cov19cpi}
\centering
\begin{tabular}[t]{p{0.8cm}lllllll}
\toprule
 &  & $\text{IS}_{W,b}^{1)}$ & $\text{IS}_{U}^{1)}$ & $\text{IS}_{50}^{1)}$ & $\text{IS}_{80}^{1)}$ & $\text{Dev}_{50}^{2)}$ & $\text{Dev}_{80}^{2)}$\\
\midrule
\parbox[t]{2mm}{\multirow{4}{*}{\rotatebox[origin=c]{90}{\parbox{2cm}{\centering Fall,\\Current}}}} & IMF & \textbf{0.114} & \textbf{0.721} & \textbf{0.559} & \textbf{0.883} & \textbf{-6} & \textbf{-16}\\
 & AR & 0.335 & 2.236 & 1.482 & 2.989 & -22 & -20\\
 & BVAR & 0.355 & 2.358 & 1.588 & 3.129 & -26 & -20\\
 & Direct: BVAR & 0.364 & 2.417 & 1.628 & 3.207 & -34 & -24\\
\addlinespace
\parbox[t]{2mm}{\multirow{4}{*}{\rotatebox[origin=c]{90}{\parbox{2cm}{\centering Spring,\\Current}}}} & IMF & 0.794 & 5.262 & 3.573 & 6.95 & -22 & -40\\
 & AR & 0.662 & 4.32 & 3.069 & 5.571 & -18 & -32\\
 & BVAR & 0.659 & 4.284 & 3.073 & 5.496 & -22 & -24\\
 & Direct: BVAR & \textbf{0.599} & \textbf{3.846} & \textbf{2.855} & \textbf{4.837} & \textbf{2} & \textbf{4}\\
\addlinespace
\parbox[t]{2mm}{\multirow{4}{*}{\rotatebox[origin=c]{90}{\parbox{2cm}{\centering Fall,\\Next}}}} & IMF & 1.739 & 11.463 & 7.9 & 15.026 & -38 & -44\\
 & AR & 1.277 & 8.077 & 6.258 & 9.897 & -26 & -32\\
 & BVAR & 1.25 & 7.851 & 6.198 & 9.504 & -26 & -20\\
 & Direct: BVAR & \textbf{1.183} & \textbf{7.331} & \textbf{6.003} & \textbf{8.659} & \textbf{-2} & \textbf{-4}\\
\addlinespace
\parbox[t]{2mm}{\multirow{4}{*}{\rotatebox[origin=c]{90}{\parbox{2cm}{\centering Spring,\\Next}}}} & IMF & 2.236 & 14.817 & 10.059 & 19.575 & -50 & -56\\
 & AR & \textbf{1.372} & \textbf{8.499} & \textbf{6.961} & 10.038 & \textbf{-6} & -16\\
 & BVAR & 1.532 & 9.608 & 7.62 & 11.596 & -22 & -28\\
 & Direct: BVAR & 1.422 & 8.675 & 7.391 & \textbf{9.959} & -14 & \textbf{0}\\
\bottomrule
\multicolumn{8}{l}{\parbox{0.81\linewidth}{\scriptsize
	${1)}$ Different Versions of the Interval Score. $\text{IS}_{50}$ and $\text{IS}_{80}$ represent the interval score for the $50\%$ and $80\%$ intervals, respectively. $\text{IS}_{W,b}$ represents the weighted sum of interval scores from quantile forecasts at base levels from main analysis $\tau \in \{0.5, 0.8\}$. $\text{IS}_{U}$ represents a simple equally-weighted average of interval scores at the same levels.}}\\
	\multicolumn{8}{l}{\parbox{0.81\linewidth}{\scriptsize
	${2)}$ Coverage deviation at nominal interval coverage levels $50\%$ and $80\%$. Negative values represent cases where the respective intervals covered fewer observations than the nominal level would indicate, and vice versa for positive values.}}
\end{tabular}
\end{table}

\newpage
\subsubsection{Illustrative Results}
\begin{figure}[h!]
    \centering
    \includegraphics[width=\linewidth]{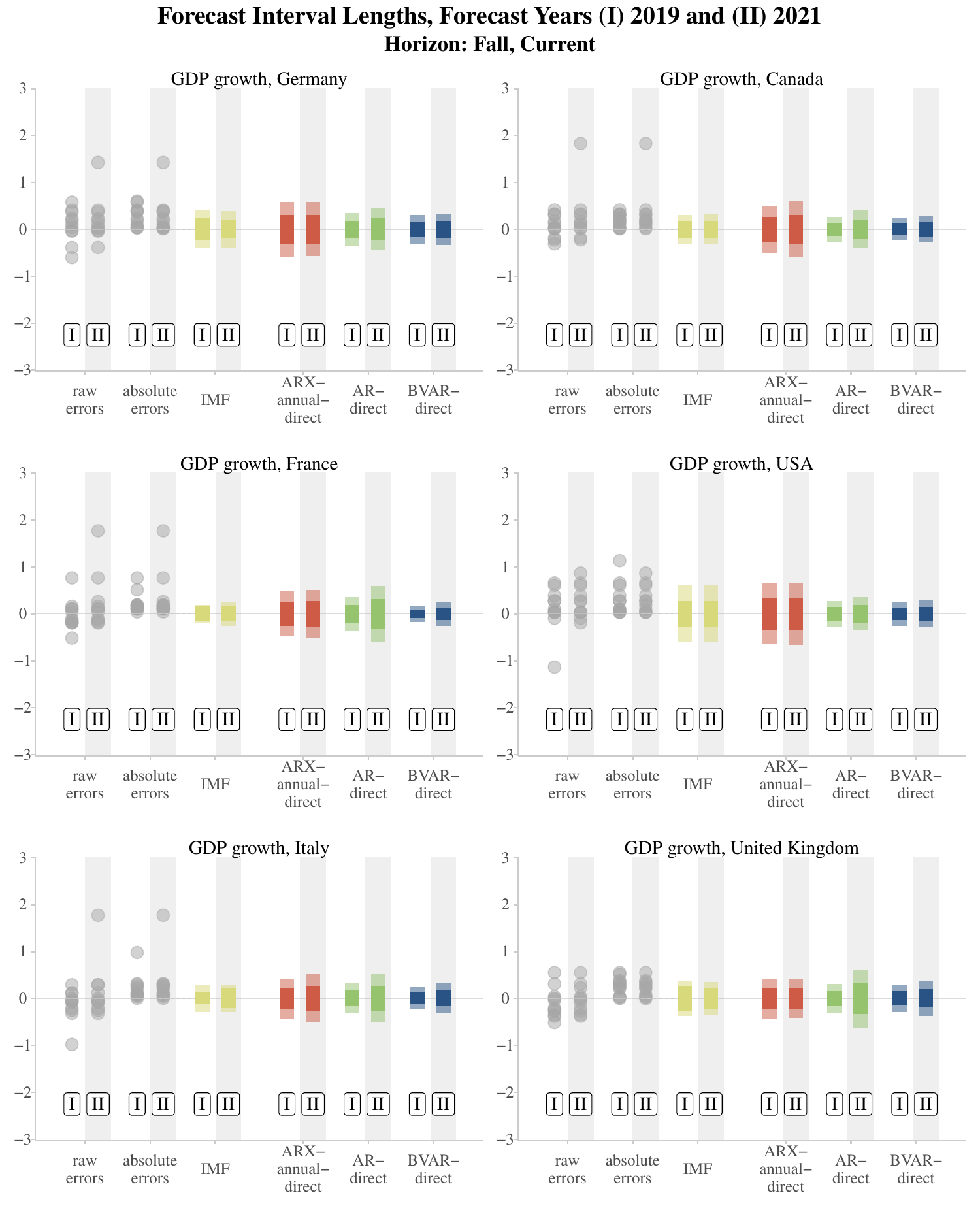}
    \caption{Raw and absolute IMF point forecast errors, and forecast intervals from different sources for the "Fall, Current" horizon, for forecast years (I) 2019 and (II) 2021, forecasting GDP growth. `IMF' forecast intervals are based on absolute IMF point forecast errors, and (I) does \textit{not} yet include the forecast error for 2020, while (II) does include it. ARX-annual is fit on annual IMF truth data and includes the IMF forecast as an additional predictor. BVAR-direct and AR-direct are fit separately on quarterly data and are shown as additional comparison. Japan is omitted, as it is not part of the evaluation set from 2021 onwards.}
    \label{fig:cov19ints_hr0gdp}
\end{figure}
\begin{figure}
    \centering
    \includegraphics[width=\linewidth]{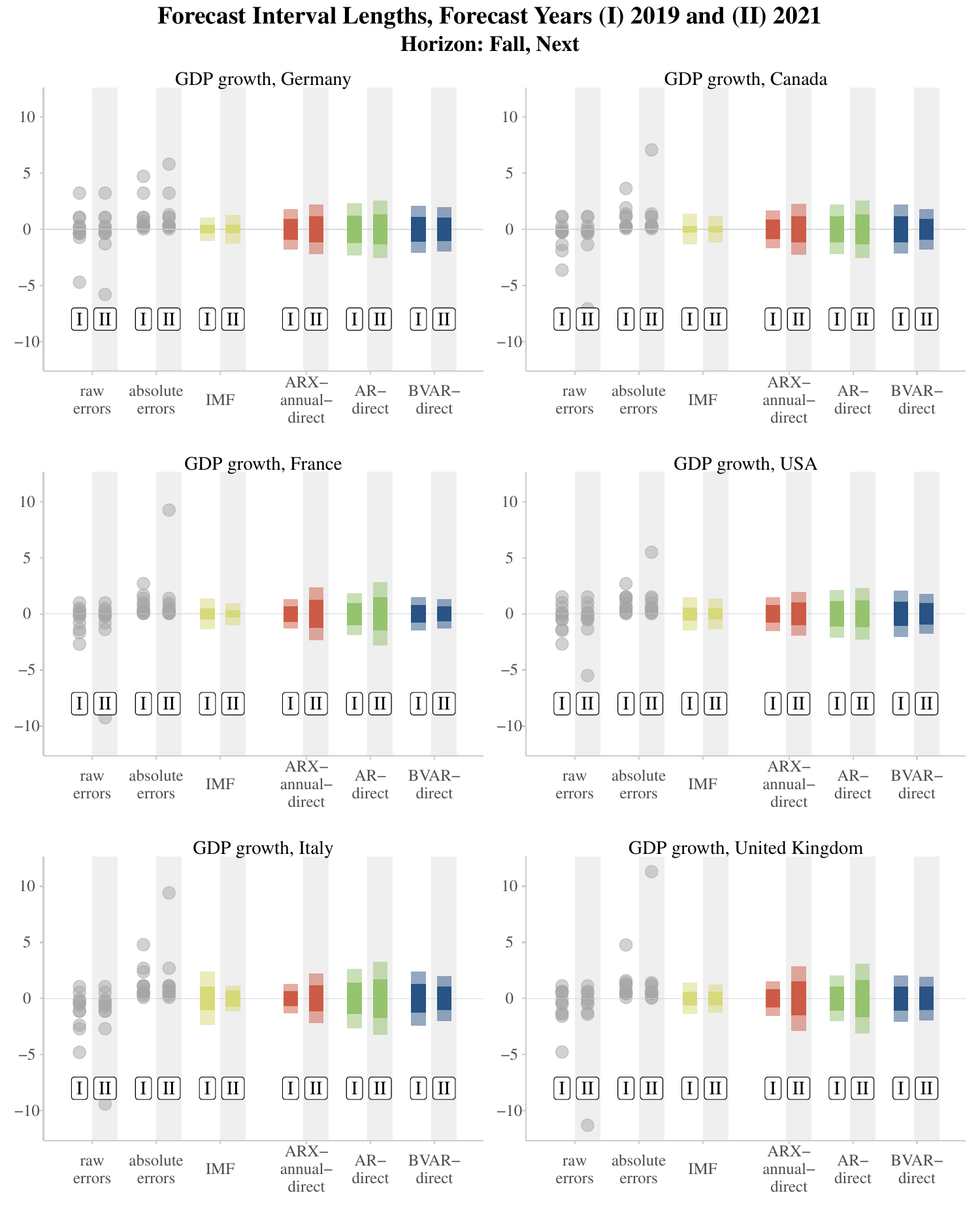}
    \caption{Same as Figure \ref{fig:cov19ints_hr0gdp}, but for forecast horizon "Fall, Next". Note that the y-axis range has been increased to accommodate higher values.}
    \label{fig:cov19ints_hr1gdp}
\end{figure}

\begin{figure}
    \centering
    \includegraphics[width=\linewidth]{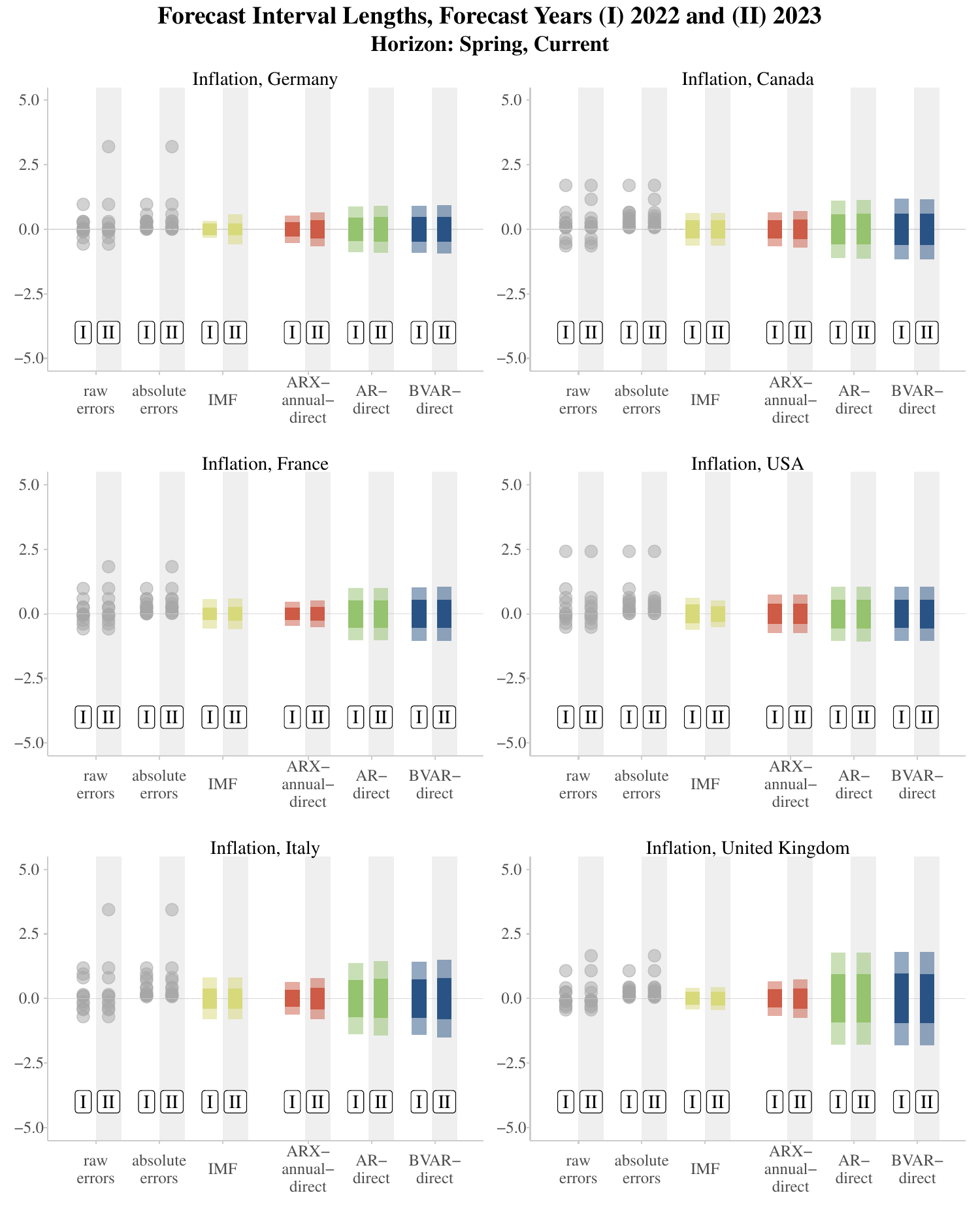}
    \caption{Same as Figure \ref{fig:cov19ints_hr0gdp}, but for inflation and forecast horizon "Spring, Current", and forecast years (I) 2022 and (II) 2023. Note that the y-axis range has been increased to accommodate higher values.}
    \label{fig:cov19ints_hr05cpi}
\end{figure}
\end{document}